\documentclass[prd,preprint,tightenlines,floatfix,showpacs,preprintnumbers,nofootinbib,eqsecnum]{revtex4}

 \usepackage[dvips,final]{graphicx}
  \usepackage{amssymb}
   \usepackage{amsmath}
    \usepackage{amsfonts}
     \usepackage{epsfig}
      \usepackage{bm}

\usepackage[section]{placeins}

\usepackage{multirow}
\usepackage{ctable}
\usepackage{booktabs}
\usepackage{array}
\usepackage{tabularx}
\usepackage{xcolor}
\usepackage{pstricks}

\def\Pom{{\bf I\!P}}
\def\Reg{{\bf I\!R}}

\begin{document}

\vfill
\title{Single- and central-diffractive production\\ of open charm and bottom mesons at the LHC:\\ theoretical predictions and experimental capabilities}

\author{Marta {\L}uszczak}
\email{luszczak@univ.rzeszow.pl} \affiliation{University of Rzesz\'ow, PL-35-959 Rzesz\'ow, Poland}

\author{Rafa{\l} Maciu{\l}a}
\email{rafal.maciula@ifj.edu.pl} \affiliation{Institute of Nuclear Physics PAN, PL-31-342 Cracow, Poland}

\author{Antoni Szczurek\footnote{also at University of Rzesz\'ow, PL-35-959 Rzesz\'ow, Poland}}
\email{antoni.szczurek@ifj.edu.pl} \affiliation{Institute of Nuclear Physics PAN, PL-31-342 Cracow, Poland}

\date{\today}

\begin{abstract}
We discuss diffractive production of open charm and bottom mesons at the LHC.
The differential cross sections for single- and central-diffractive mechanisms for $c\bar c$ and $b\bar b$ pair production are calculated in the framework of the Ingelman-Schlein model corrected for absorption effects. 
In this approach one assumes that the pomeron has a well defined partonic structure, and that the hard process takes
place in a pomeron-proton or proton-pomeron (single diffraction) or pomeron-pomeron (central diffraction) processes. Here, leading-order gluon-gluon fusion and quark-antiquark
anihilation partonic subprocesses are taken into consideration, which are calculated within
standard collinear approximation. Both pomeron flux factors as well as parton distributions in the
pomeron are taken from the H1 Collaboration analysis of diffractive
structure function and diffractive dijets at HERA. The extra corrections from subleading reggeon exchanges are
explicitly calculated and are also taken into consideration.
Several quark-level differential distributions are shown.
The hadronization of charm and bottom quarks is taken into account by means of fragmentation function technique.
Predictions for single- and central-diffractive production in the case of inclusive $D$ and $B$ mesons,
as well as $D\bar D$ pairs are presented, including detector acceptance of the ATLAS, CMS and LHCb Collaborations.
The experimental
aspects of possible standard and dedicated measurements are carefully discussed.
\end{abstract}

\pacs{13.87.Ce,14.65.Dw}

\maketitle

\section{Introduction}
Diffractive processes were intensively studied at HERA
in $\gamma p$ and $e p$ collisions for more than a decade.
On theoretical side, somewhat enigmatically, these are processes 
with exchange of pomeron or processes with the QCD amplitude without net
color exchange. In such processes pomeron must be treated rather technically,
depending on the formulation of the approach.
Experimentally such processes are defined by special requirement(s)
on the final state. The most popular is a requirement of rapidity gap
starting from the final proton(s) on one (single-diffrative process) or both
(central-diffractive process) sides. The size of the gap is essentially
experimental observable but it is not easy to calculate theoretically. 
Several processes with different final states were studied
at HERA, such as dijet, charm production, etc. 
The H1 Collaboration has found a set of so-called
diffractive parton distributions in the proton inspired by 
the Ingelman-Schlein model \cite{IS}. In this fit both pomeron 
and reggeon contributions were included. We wish to emphasize that 
there is no common consensus as far as a model of diffractive production
is considered. However, these open problems go beyond the scope of 
the present paper and will be not discussed here. 

One can gain a better understanding of the mechanism of the diffractive
production by going from photon-proton to proton-proton or
proton-antiproton scattering.
There, however, some new elements related to nonperturbative interaction
between protons show up, such as absorption effects.
So far only some selected diffractive processes were discussed in 
the literature such as diffractive production of dijets \cite{Kramer}, 
production of $W$ \cite{Collins} and $Z$ \cite{CSS09} bosons, 
production of $W^+ W^-$ pairs \cite{LSR} or production of 
$c \bar c$ \cite{LMS2011}. 
The latter was done there only for illustration of the general situation
at the parton level.
The cross section for diffractive processes are in general rather
small (e.g. the single-diffractive processes are of the order of 
a few percent compared to inclusive cross sections).
In order to measure rapidity gap(s) the luminosity cannot be big
to avoid so-called pile-ups \cite{Royon}. 
All this causes that for some interesting processes, as for instance
$W$ or $Z^0$ production, the statistics is rather poor and the cross 
section is difficult to measure.
Since the cross section for inclusive production of charm
is very large at the LHC \cite{Maciula:2013wg}, one could expect that also single- 
and central- diffractive charm production could be measured with 
relatively good precision. This is therefore a process one could
use for testing theoretical models. The same shall be true
for diffractive bottom production.

It is the aim of this paper to present predictions including
our knowledge about diffractive parton distributions from HERA
and taking into account absorption effects, specific for proton-proton
collisions. We shall include both pomeron and reggeon contributions. 
In addition, we shall include hadronization
of $c$ or $b$ quarks/antiquarks to open charmed and bottom mesons, respectively.
Finally we shall present our predictions for experiments at the LHC.
We hope that our predictions will be verified at the LHC in a near future.
\section{Theoretical framework}
\subsection{A sketch of formalism}
\label{formalism}

\begin{figure}[!ht]
\begin{flushleft}
{\scriptsize a) \hskip+77mm b)}
\end{flushleft} \vskip-5mm
\includegraphics[height=.13\textheight]{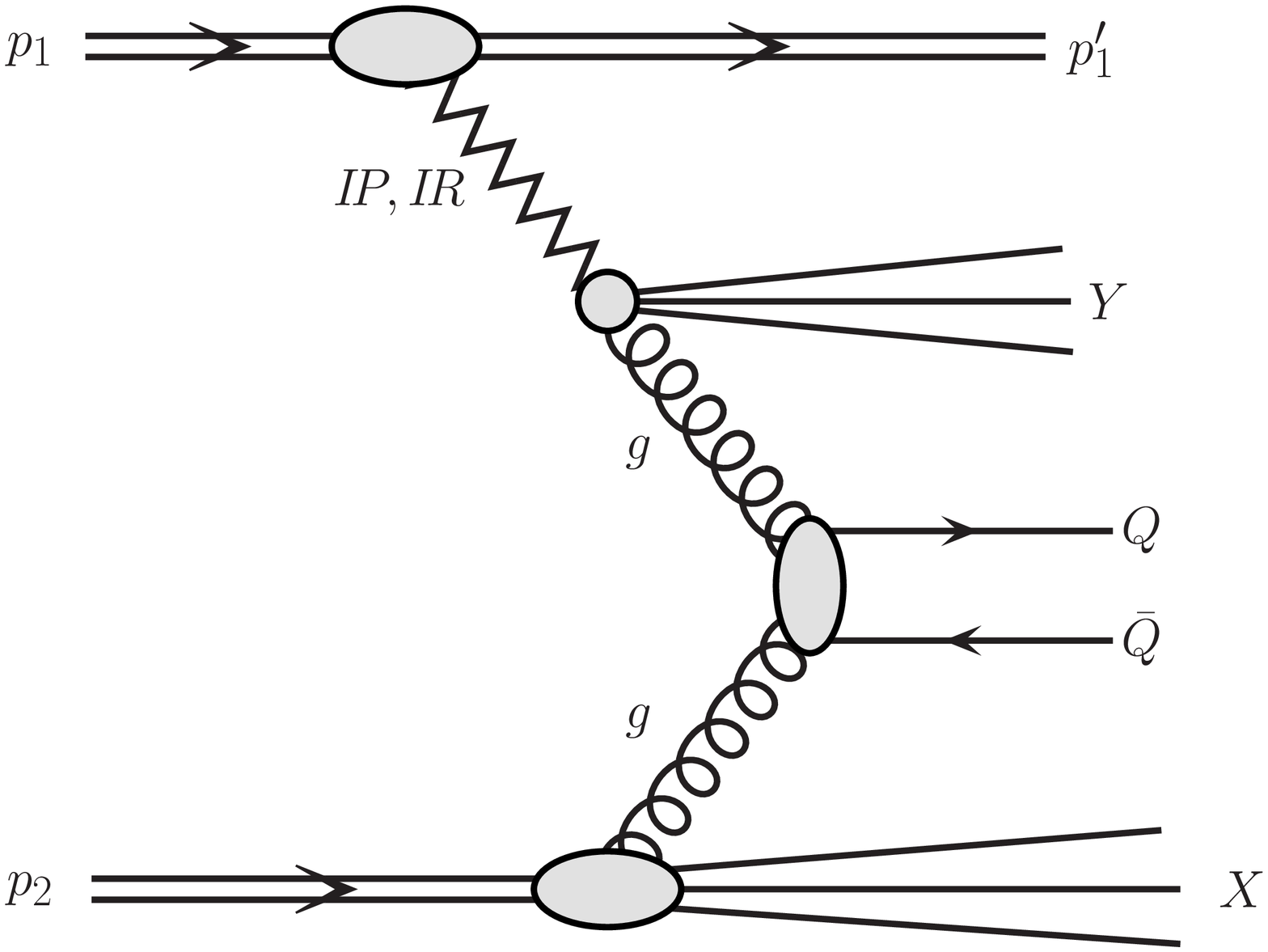}\hskip-1mm
\includegraphics[height=.13\textheight]{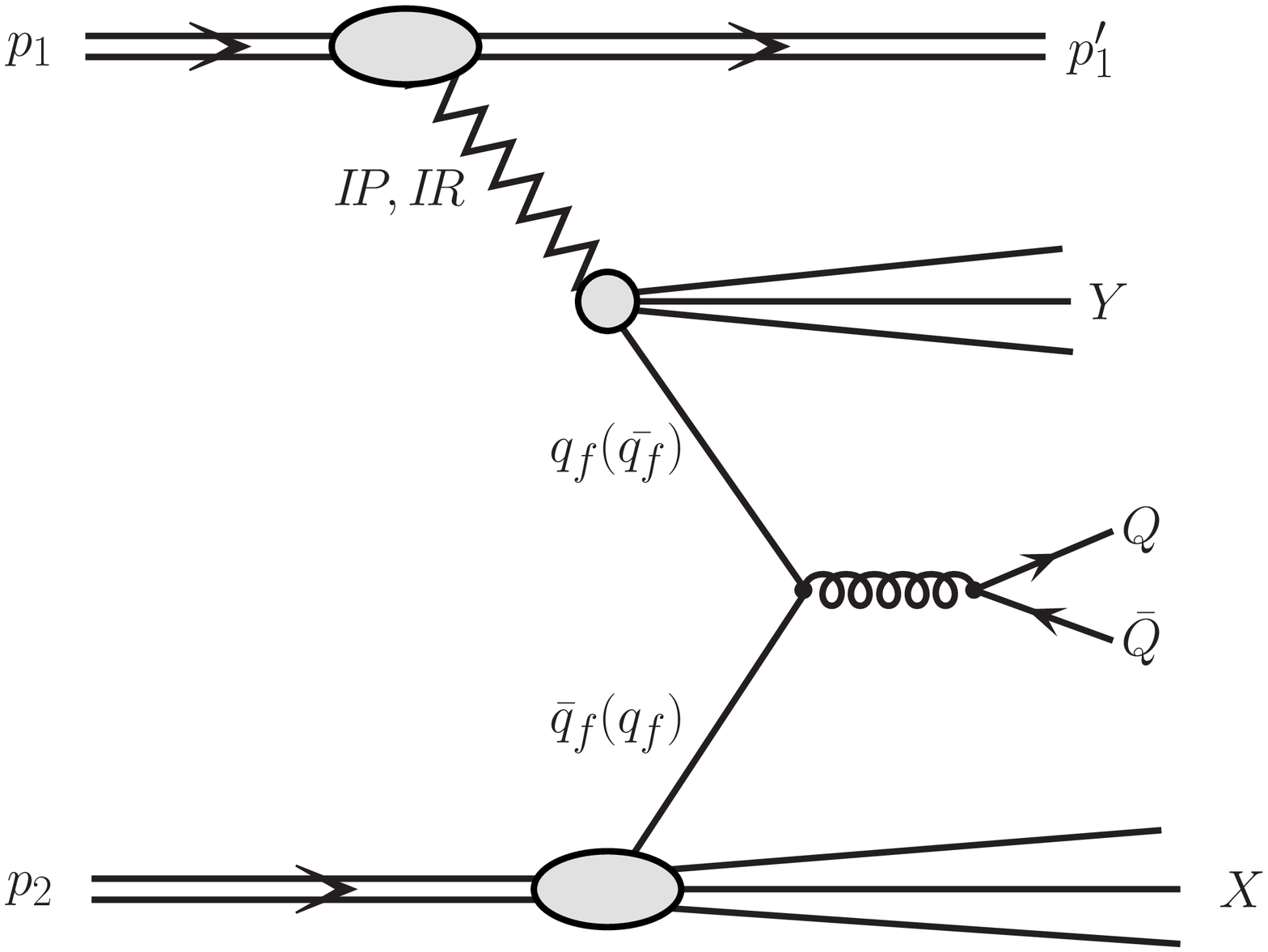}\hskip-1mm
\includegraphics[height=.13\textheight]{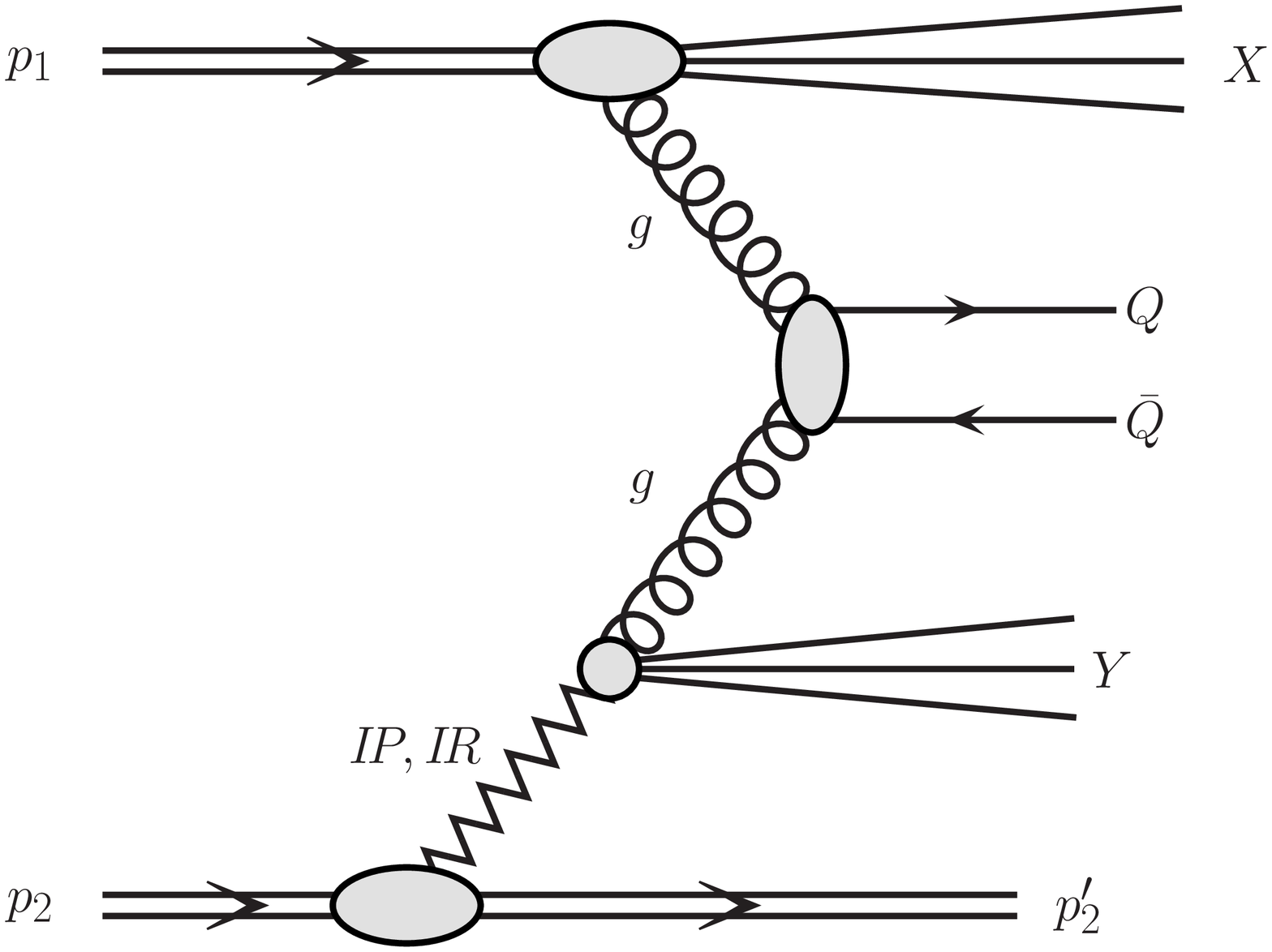}\hskip-1mm
\includegraphics[height=.13\textheight]{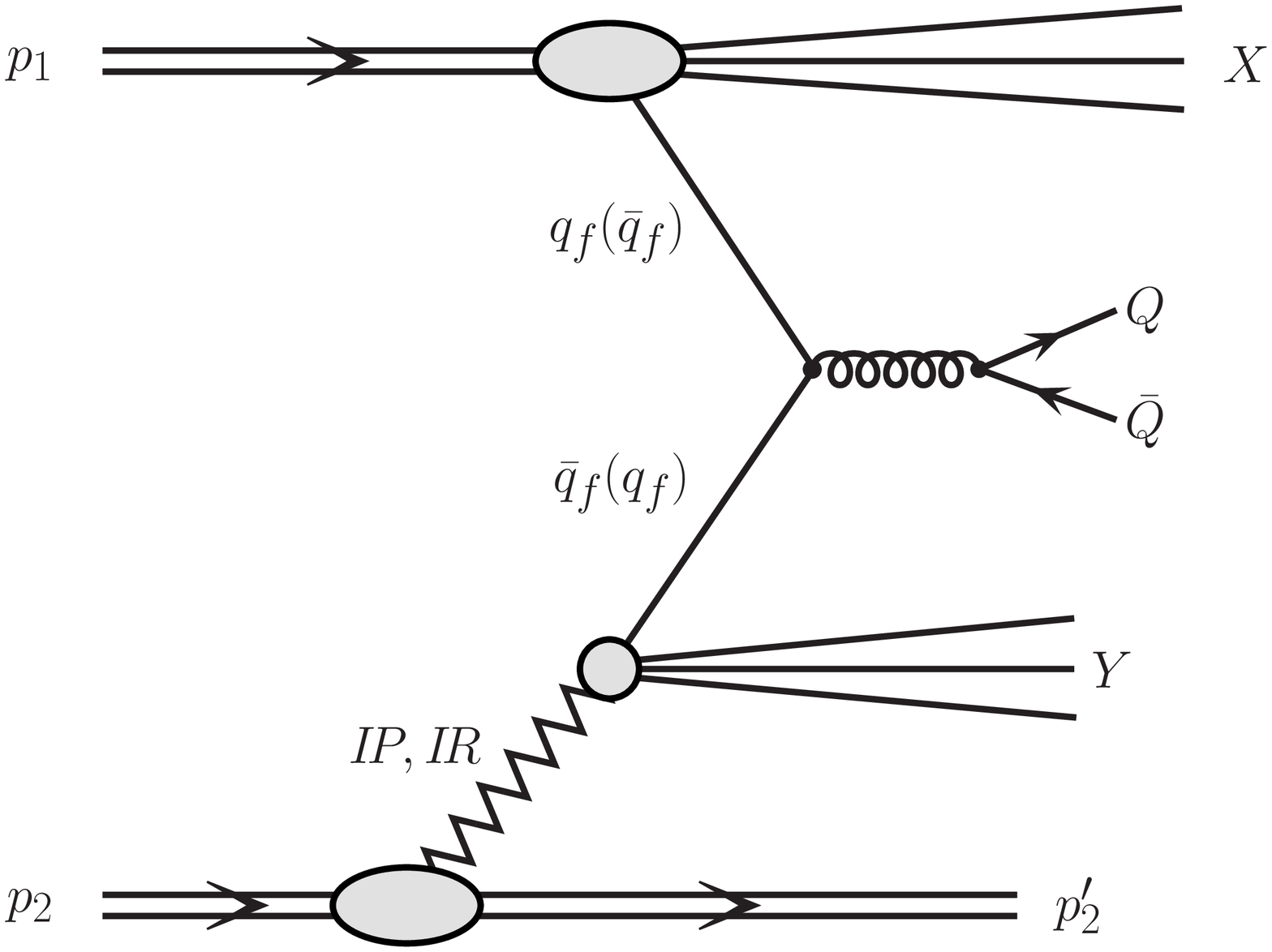}
\caption{
The mechanisms of single-diffractive production of heavy quarks.
}
 \label{fig:1}
\end{figure}

\begin{figure}[!ht]
\includegraphics[height=.13\textheight]{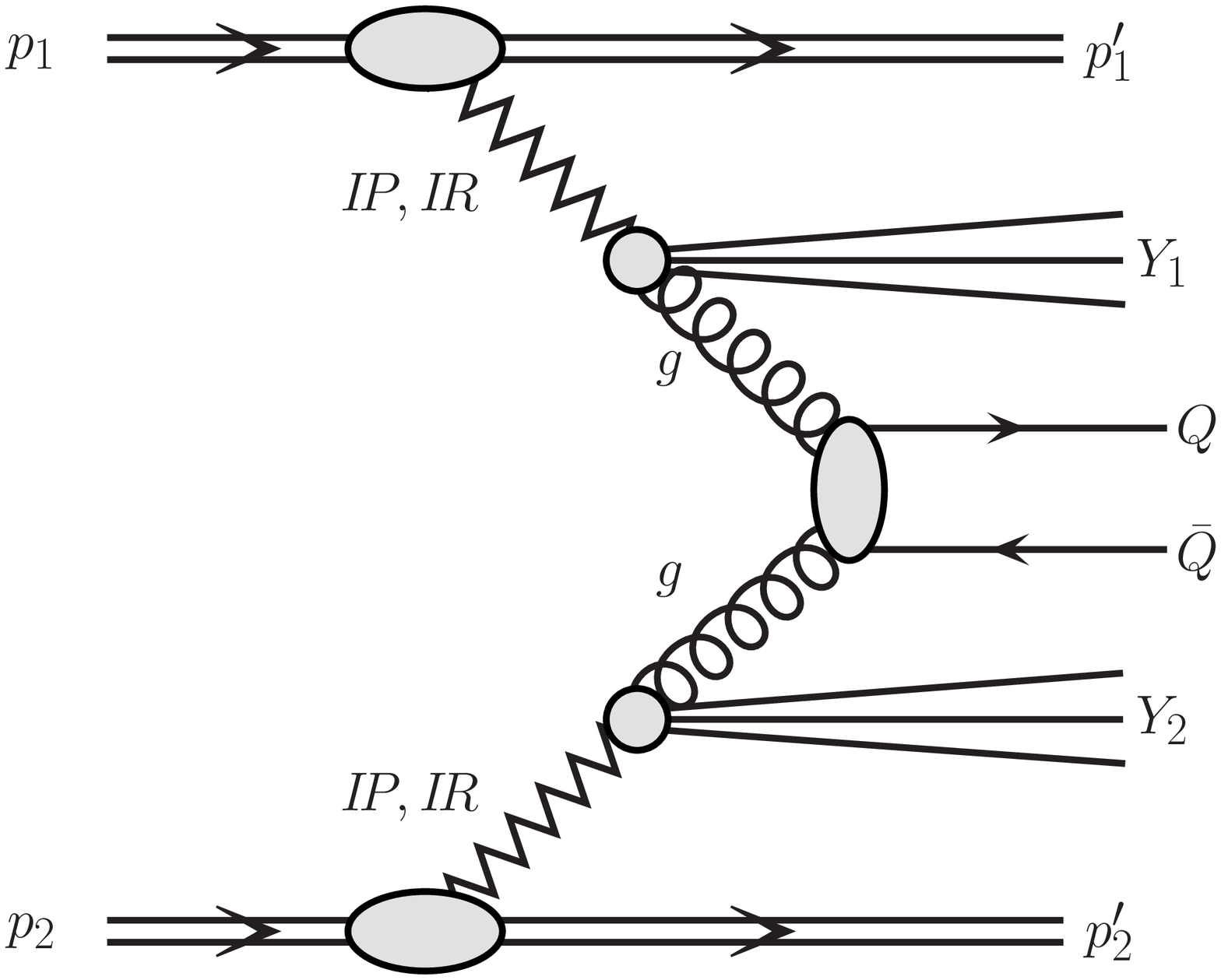}
\includegraphics[height=.13\textheight]{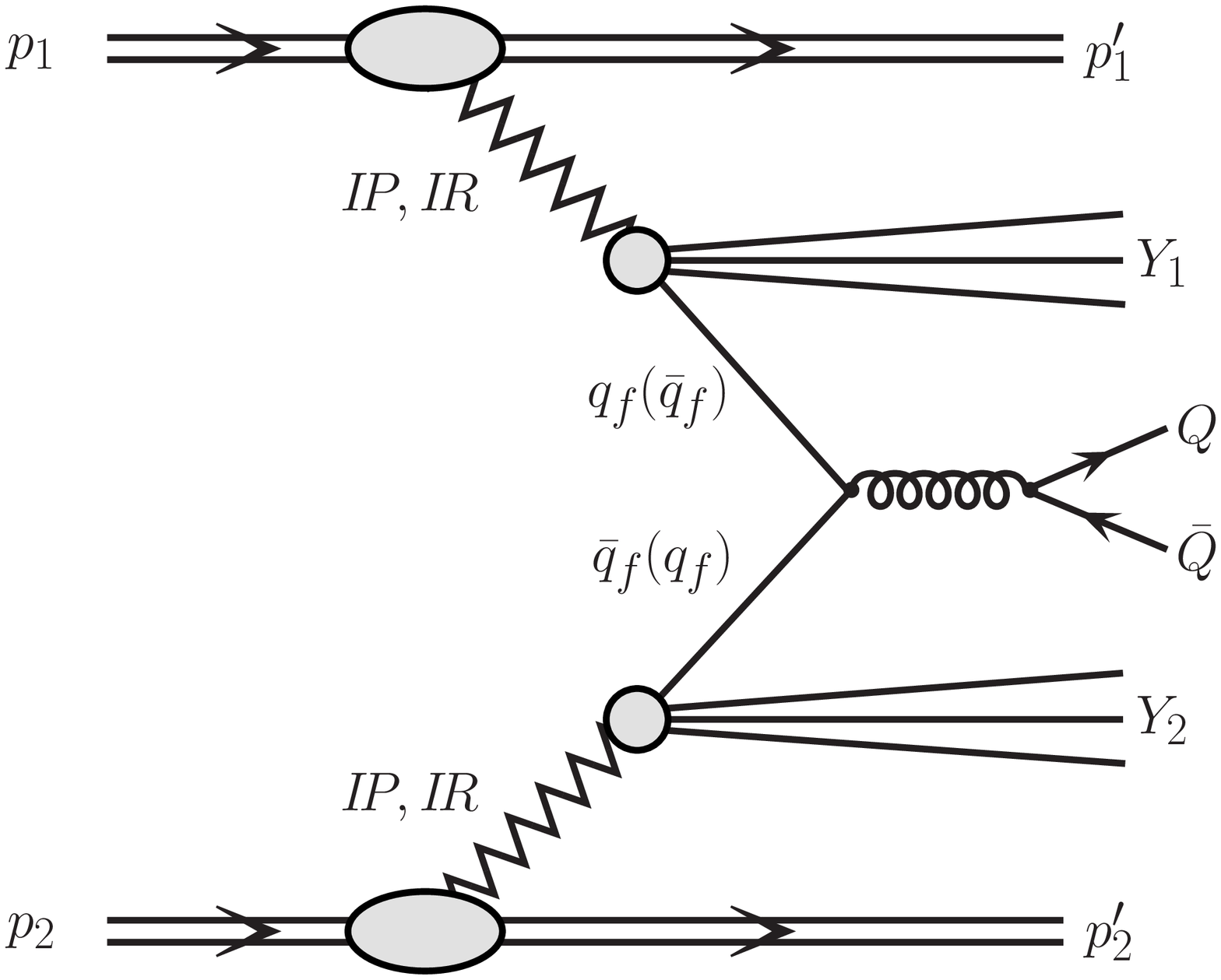}
\caption{
The mechanisms of central-diffractive production of heavy quarks. 
}
\label{fig:2}
\end{figure}

The mechanisms of the diffractive production 
of heavy quarks ($c \bar c$, $b \bar b$) discussed here are shown in 
Figs.~\ref{fig:1} and ~\ref{fig:2}. Both, leading-order (LO)
gg-fusion and $q \bar q$-anihilation partonic subprocesses
are taken into account in the calculations.

In the following we apply the Ingelman-Schlein approach \cite{IS}. 
In this approach one assumes that the pomeron has a
well defined partonic structure, and that the hard process
takes place in a pomeron--proton or proton--pomeron (single diffraction) 
or pomeron--pomeron (central diffraction) processes.
In this approach corresponding differential cross sections can be
written as
\begin{eqnarray}
{d \sigma_{SD^{(1)}} \over dy_{1} dy_{2} dp_{t}^2} &=& {1 \over 16 \pi^2 \hat{s}^2} \times \Big[ |{\cal M}_{g g \rightarrow Q \bar{Q}}|^2 \cdot
 x_1 g^{D}(x_1,\mu^2) x_2 g(x_2,\mu^2) \nonumber \\
&+& |{\cal M}_{q \bar{q} \rightarrow Q \bar{Q}}|^2 \cdot \Big( x_1 q^{D}(x_1,\mu^2) x_2 \bar{q}(x_2,\mu^2)
+ x_1 \bar{q}^{D}(x_1,\mu^2) x_2 q(x_2,\mu^2) \Big) \Big] \; ,
\label{SD1}
\end{eqnarray}
\begin{eqnarray}
{d \sigma_{SD^{(2)}} \over dy_{1} dy_{2} dp_{t}^2} &=& {1 \over 16 \pi^2 \hat{s}^2} \times \Big[ |{\cal M}_{g g \rightarrow Q \bar{Q}}|^2 \cdot
 x_1 g(x_1,\mu^2) x_2 g^{D}(x_2,\mu^2) \nonumber \\
&+& |{\cal M}_{q \bar{q} \rightarrow Q \bar{Q}}|^2 \cdot \Big( x_1 q (x_1,\mu^2) x_2 \bar{q}^{D}(x_2,\mu^2) 
+ x_1 \bar{q}(x_1,\mu^2) x_2 q^{D}(x_2,\mu^2) \Big) \Big] \; ,
\label{SD2}
\end{eqnarray}
\begin{eqnarray}
{d \sigma_{CD} \over dy_{1} dy_{2} dp_{t}^2} &=&  {1 \over 16 \pi^2 \hat{s}^2} \times \Big[ |{\cal M}_{g g \rightarrow Q \bar{Q}}|^2 \cdot  x_1 g^{D}(x_1,\mu^2) x_2 g^{D}(x_2,\mu^2) \\
&+& |{\cal M}_{q \bar{q} \rightarrow Q \bar{Q}}|^2 \cdot \Big( x_1 q^{D}(x_1,\mu^2) x_2 \bar q^{D}(x_2,\mu^2) + x_1 \bar q^{D}(x_1,\mu^2)
\, x_2  q^{D}(x_2,\mu^2) \Big) \,\Big ] \; , \nonumber  
\label{DD}
\end{eqnarray}
for single-diffractive (SD) and central-diffractive (CD) production, 
respectively.

The diffractive distribution function (diffractive PDF) can be obtained by a convolution of the flux of pomerons
$f_\Pom(x_\Pom)$ in the proton and the parton distribution in the pomeron,
e.g. $g_{\Pom}(\beta, \mu^2)$ for gluons:
\begin{eqnarray}
g^D(x,\mu^2) = \int d x_\Pom d\beta \, \delta(x-x_\Pom \beta) 
g_{\Pom} (\beta,\mu^2) \, f_\Pom(x_\Pom) \, 
= \int_x^1 {d x_\Pom \over x_\Pom} \, f_\Pom(x_\Pom)  
g_{\Pom}({x \over x_\Pom}, \mu^2) \, . \nonumber \\
\end{eqnarray}
The flux of pomerons $f_\Pom(x_\Pom)$ enters in the form integrated over 
four--momentum transfer 
\begin{eqnarray}
f_\Pom(x_\Pom) = \int_{t_{min}}^{t_{max}} dt \, f(x_\Pom,t) \, ,
\label{flux_of_Pom}
\end{eqnarray}
with $t_{min}, t_{max}$ being kinematic boundaries.

Both pomeron flux factors $f_{\Pom}(x_{\Pom},t)$ as well 
as parton distributions in the pomeron were taken from 
the H1 Collaboration analysis of diffractive structure function
and diffractive dijets at HERA \cite{H1}. 
In the following calculation standard collinear MSTW08LO parton distributions are
used \cite{MSTW08}. 
The renormalization scale in $\alpha_s$ and factorization scale for the
diffractive PDFs are taken to be equal to heavy quark transverse mass
$\mu = m_t$ as a default and $\mu = \hat s$ for illustration of related
uncertainty. The heavy quark mass in the calculation
is set to 1.5 and 4.75 GeV for charm and bottom, respectively.

\subsection{Results for diffractive $Q \bar Q$ pair production}

Let us start presentation of our results for diffraction mechanisms.
In the present analysis we consider both pomeron and subleading reggeon
contributions. In the H1 Collaboration analysis the pion structure 
function was used for the subleading reggeons and the corresponding 
flux was fitted to the diffractive DIS data.
The corresponding diffractive parton distributions
are obtained by replacing the pomeron flux by the reggeon flux and
the parton distributions in the pomeron by their counterparts
in subleading reggeon \cite{H1}.

\begin{figure}[!h]
\begin{minipage}{0.47\textwidth}
 \centerline{\includegraphics[width=1.0\textwidth]{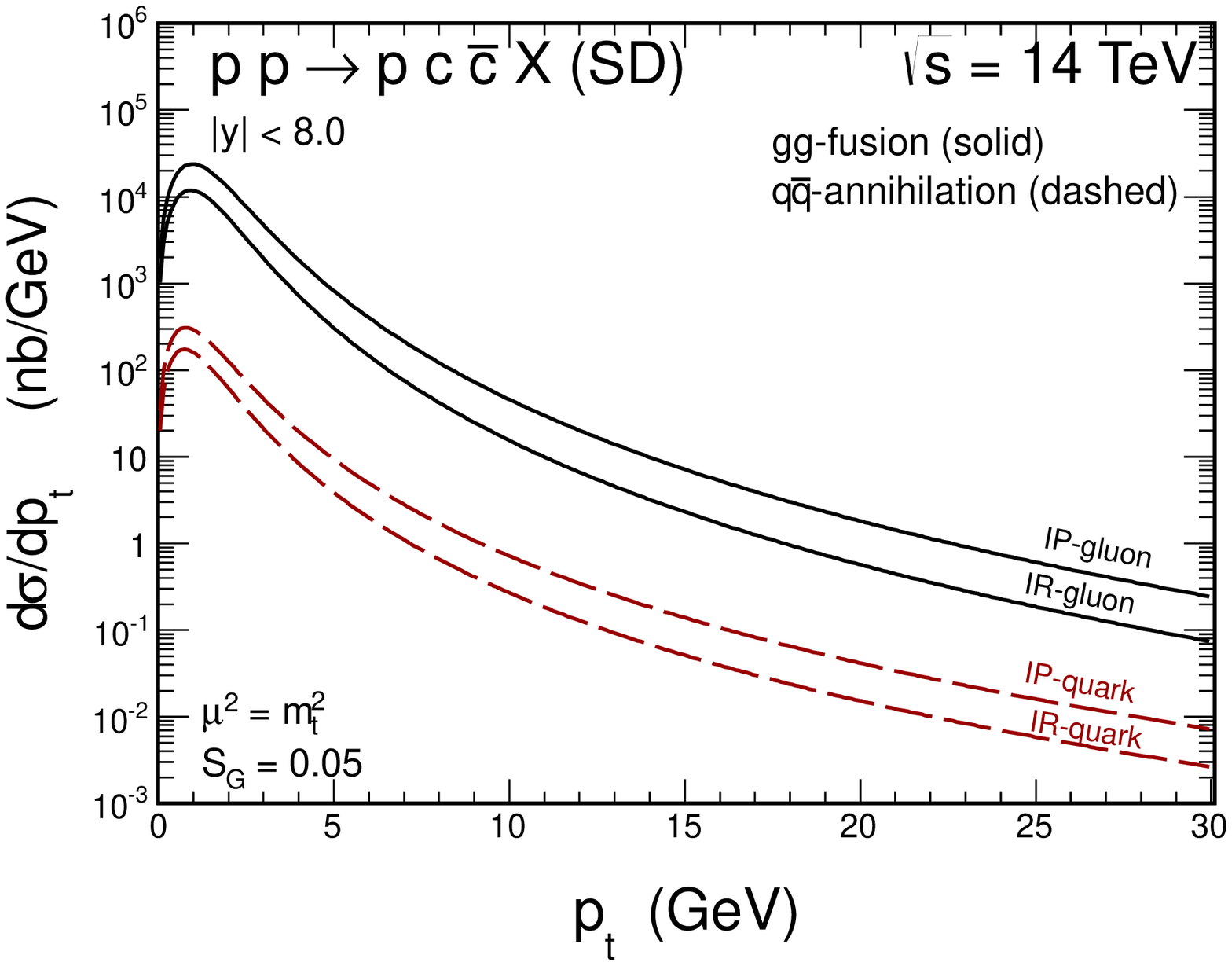}}
\end{minipage}
\hspace{0.5cm}
\begin{minipage}{0.47\textwidth}
 \centerline{\includegraphics[width=1.0\textwidth]{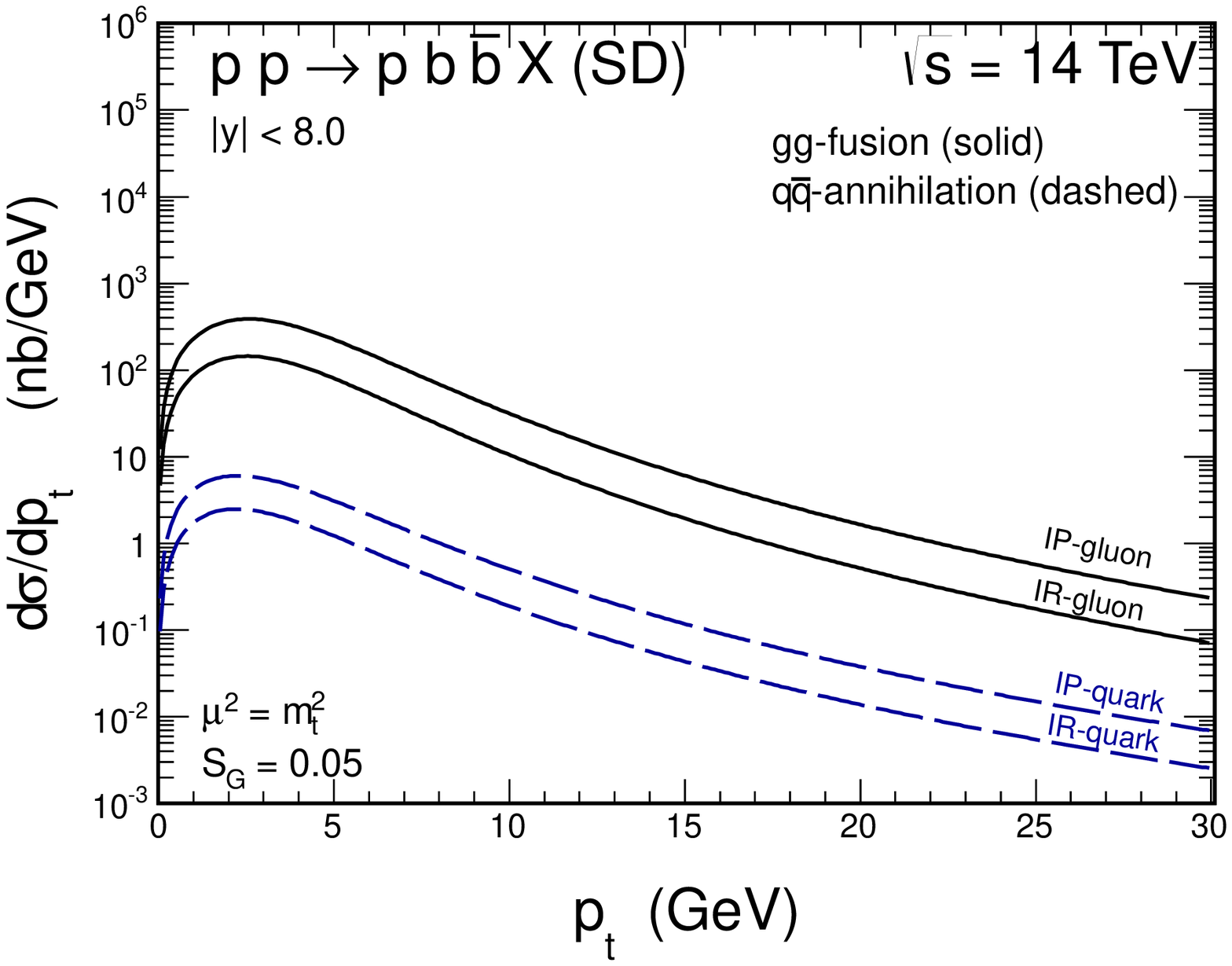}}
\end{minipage}
   \caption{
\small Transverse momentum distribution of $c$ quarks (antiquarks) (left)
and $b$ quarks (antiquarks) (right) for single-diffractive production at $\sqrt{s} = 14$ TeV.
Components of the pomeron-gluon (and gluon-pomeron), the pomeron-quark(antiquark) (and quark(antiquark)-pomeron) and the reggeon-gluon (and gluon-reggeon),
the reggeon-quark(antiquark) (and quark(antiquark)-reggeon) mechanisms are shown separately.
}
 \label{fig:pt}
\end{figure}

In Fig.~\ref{fig:pt} we show the transverse momentum distribution of $c$ quarks (antiquarks)
and $b$ quarks (antiquarks) for single-diffractive production at $\sqrt{s} = 14$ TeV. 
Contributions of the pomeron-gluon (and gluon-pomeron), the pomeron-quark(antiquark) (and quark(antiquark)-pomeron) and the reggeon-gluon (and gluon-reggeon),
the reggeon-quark(antiquark) (and quark(antiquark)-reggeon) mechanisms are shown separately.
Components of the pomeron-gluon (and gluon-pomeron)
are almost two orders of magnitude larger than
the pomeron-quark(antiquark) and quark(antiquark)-pomeron.
The estimated reggeon contribution is
of similar size as the leading pomeron contribution, but still slightly smaller.

The calculation done assumes Regge factorization, which is known
to be violated in hadron-hadron collisions.
It is known that soft interactions lead to an extra production
of particles which fill in the rapidity gaps related to pomeron
exchange.

Different models of absorption corrections 
(one-, two- or three-channel approaches) 
for diffractive processes were presented in the literature.
The absorption effects for the diffractive processes were calculated e.g.
in \cite{KMR2000,Maor2009,CSS09}.
The different models give slightly different predictions.
Usually an average value of the gap survival probability
$<|S_G|^2>$ is calculated first and then the cross sections for different
processes is multiplied by this value.
We shall follow this somewhat simplified approach also here.
Numerical values of the gap survival probability can be found 
in \cite{KMR2000,Maor2009,CSS09}.
The survival probability depends on the collision energy.
It is sometimes parametrized as:
\begin{equation}
<S_G^2>(\sqrt{s}) = \frac{a}{b+\ln(\sqrt{s})} \; .
\end{equation}
The multiplicative factors are approximately $S_G$ = 0.05
for single-diffractive production 
and $S_G$ = 0.02 for central-diffractive one for the nominal LHC energy ($\sqrt{s}=$ 14 TeV).

\begin{figure}[!h]
\begin{minipage}{0.47\textwidth}
 \centerline{\includegraphics[width=1.0\textwidth]{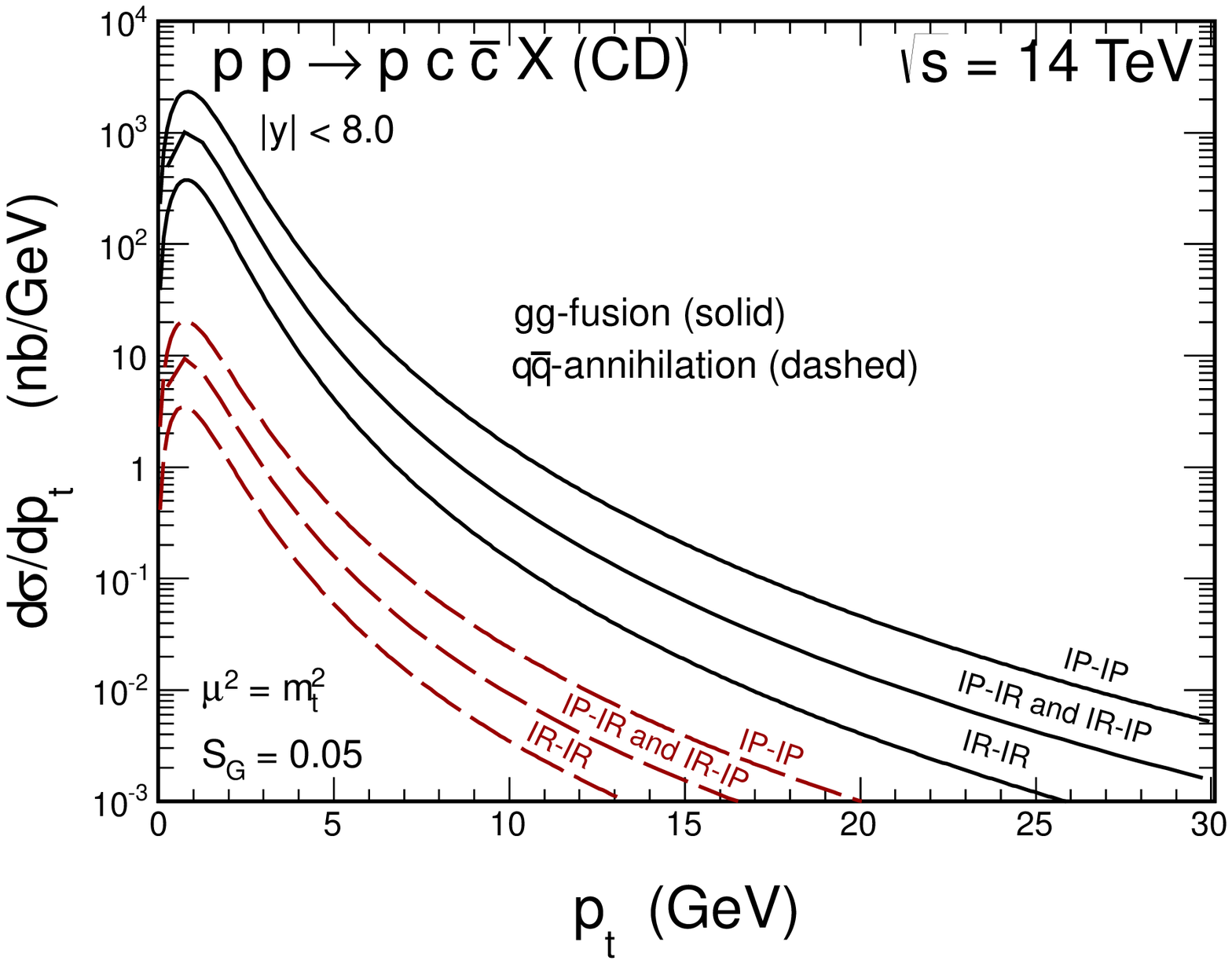}}
\end{minipage}
\hspace{0.5cm}
\begin{minipage}{0.47\textwidth}
 \centerline{\includegraphics[width=1.0\textwidth]{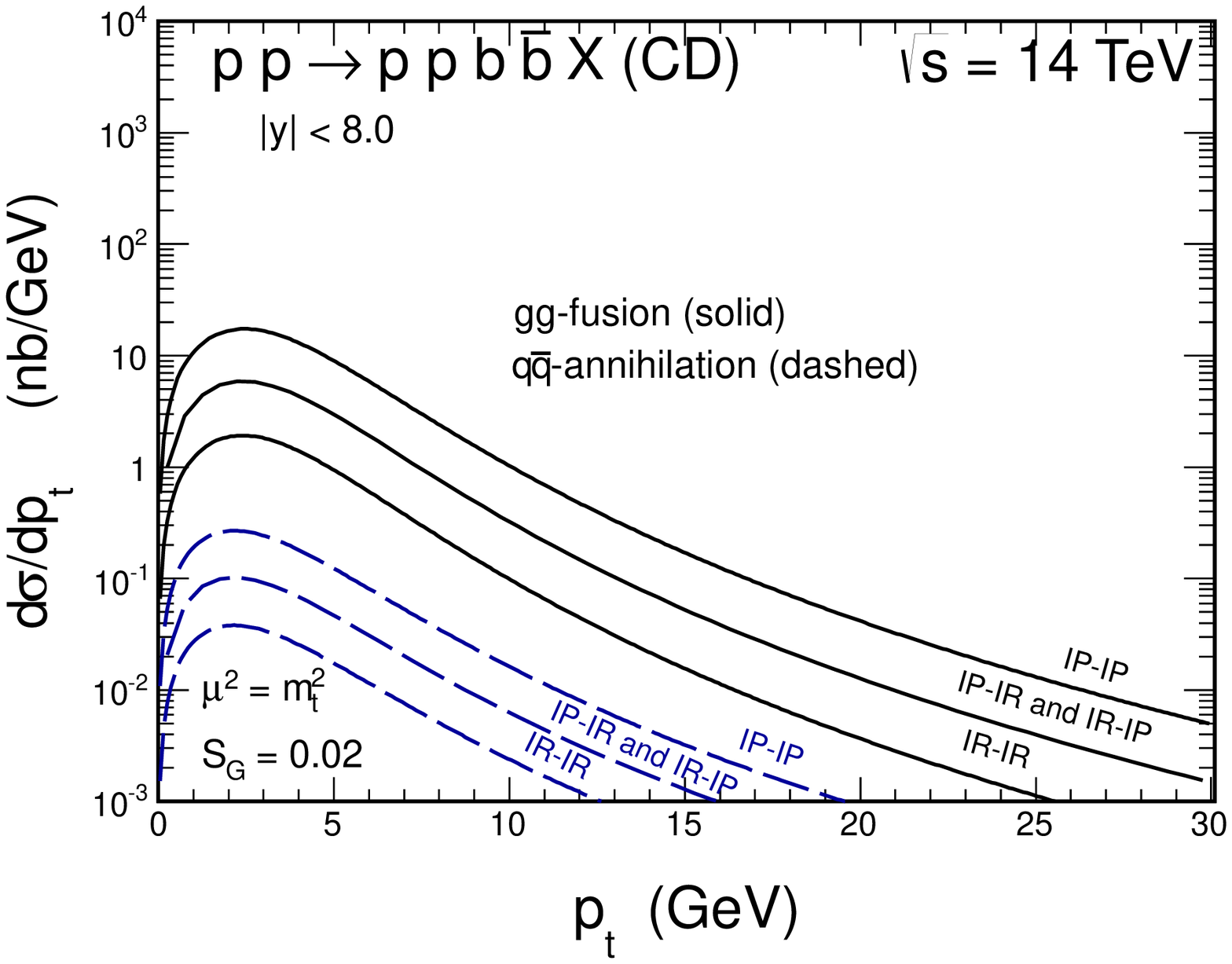}}
\end{minipage}
   \caption{
\small Transverse momentum distribution of $c$ quarks (antiquarks) (left)
and $b$ quarks (antiquarks) (right) for the central-diffractive production at $\sqrt{s} = 14$ TeV.
Components of the pomeron-pomeron, reggeon-reggeon, Pomerom-reggeon and reggeon-pomeron mechanisms are shown separately.
}
 \label{fig:pt_CD}
\end{figure}

In Fig.~\ref{fig:pt_CD} we show the transverse momentum distribution of
$c$ quarks (antiquarks)
and $b$ quarks (antiquarks) for central-diffractive production at $\sqrt{s} = 14$ TeV.
The distributions for central-dffractive component is smaller than that for the single-diffractive distributions by almost two orders of magnitude. 

\begin{figure}[!h]
\begin{minipage}{0.47\textwidth}
 \centerline{\includegraphics[width=1.0\textwidth]{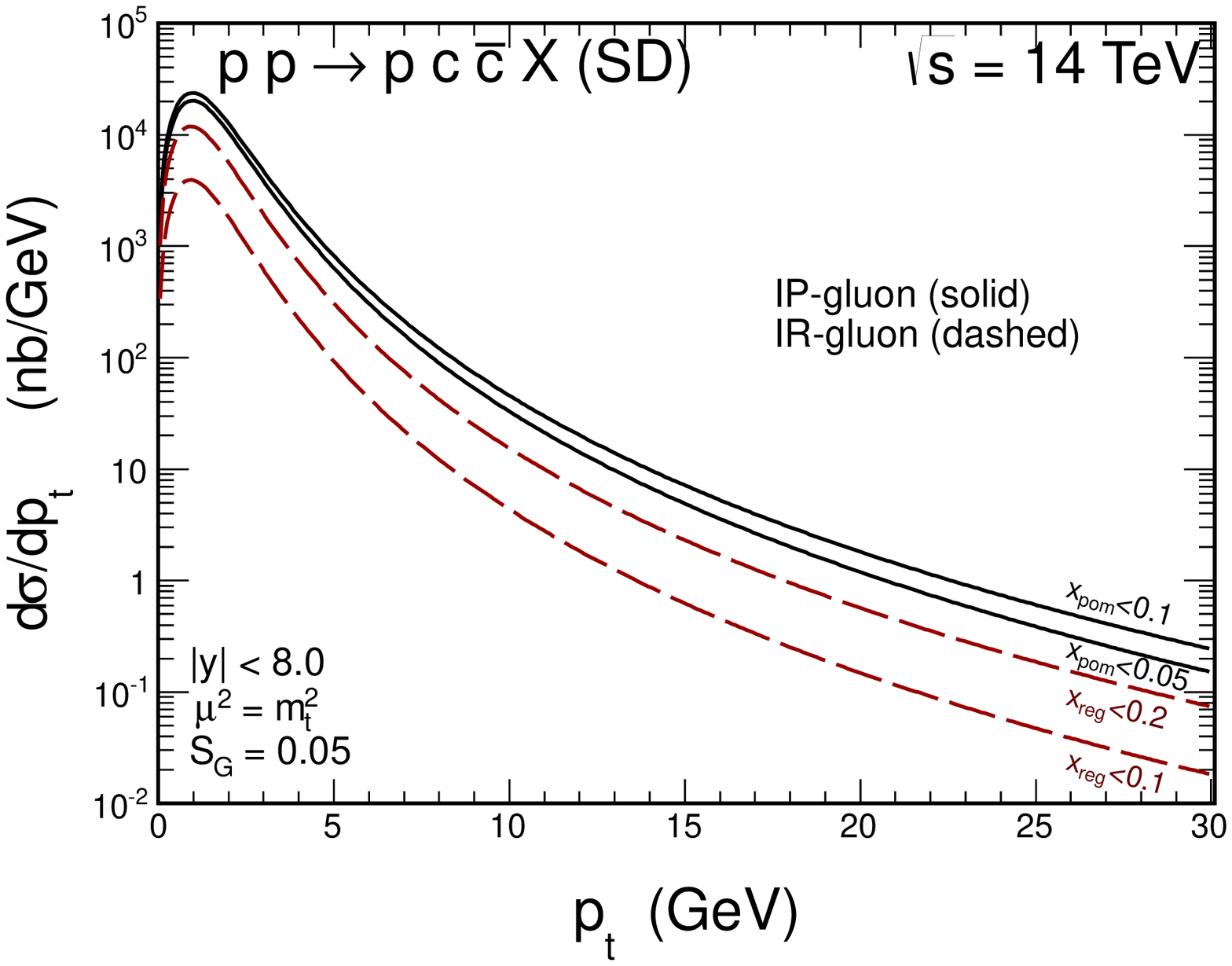}}
\end{minipage}
\hspace{0.5cm}
\begin{minipage}{0.47\textwidth}
 \centerline{\includegraphics[width=1.0\textwidth]{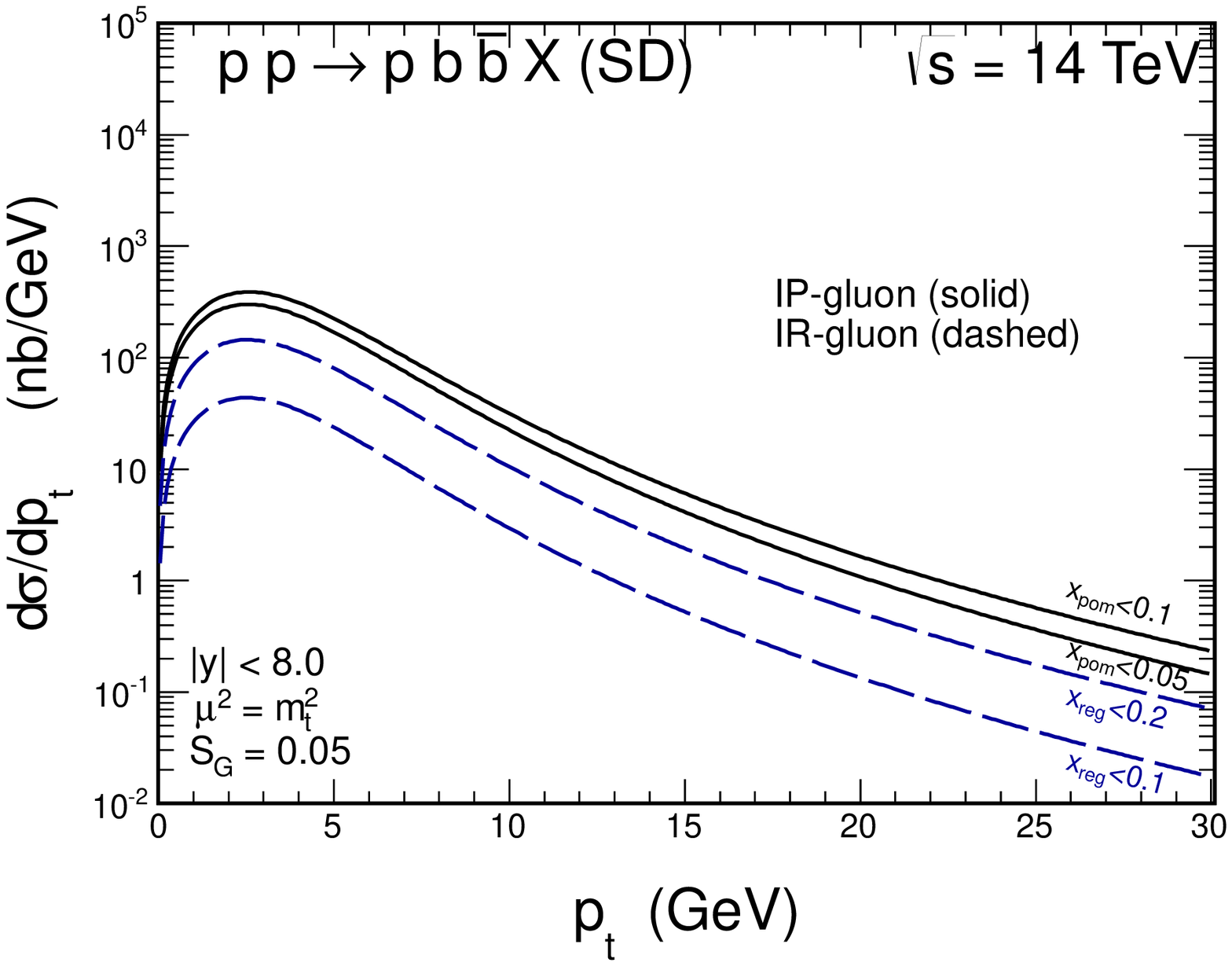}}
\end{minipage}
   \caption{
\small Transverse momentum distribution of $c$ quarks (antiquarks) (left)
and $b$ quarks (antiquarks) (right) for single-diffractive production at $\sqrt{s} = 14$ TeV for different maximal $x_\Pom$ (solid) and $x_\Reg$ (dashed). 
}
 \label{fig:pt_x}
\end{figure}

\begin{figure}[!h]
\begin{minipage}{0.47\textwidth}
 \centerline{\includegraphics[width=1.0\textwidth]{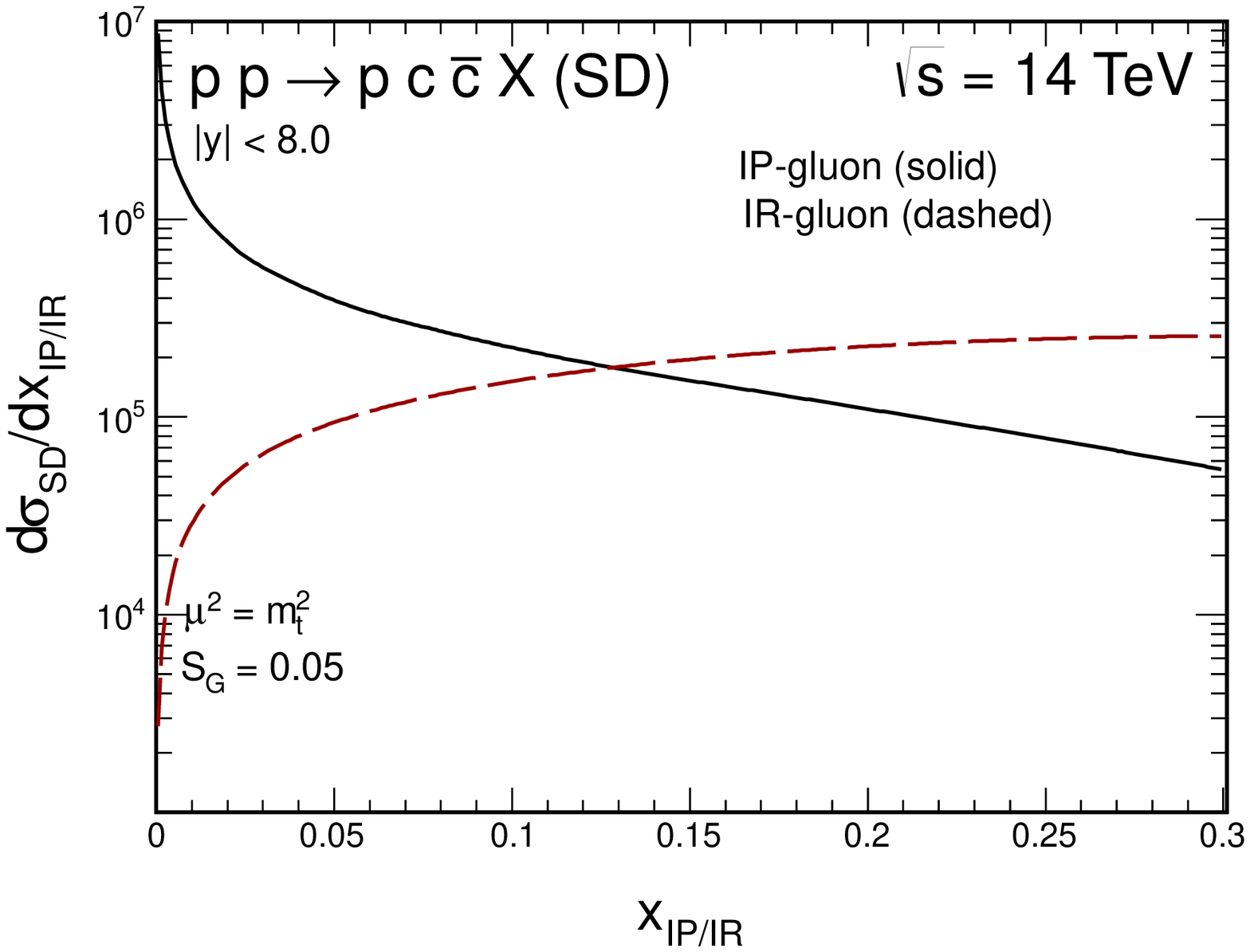}}
\end{minipage}
\hspace{0.5cm}
\begin{minipage}{0.47\textwidth}
 \centerline{\includegraphics[width=1.0\textwidth]{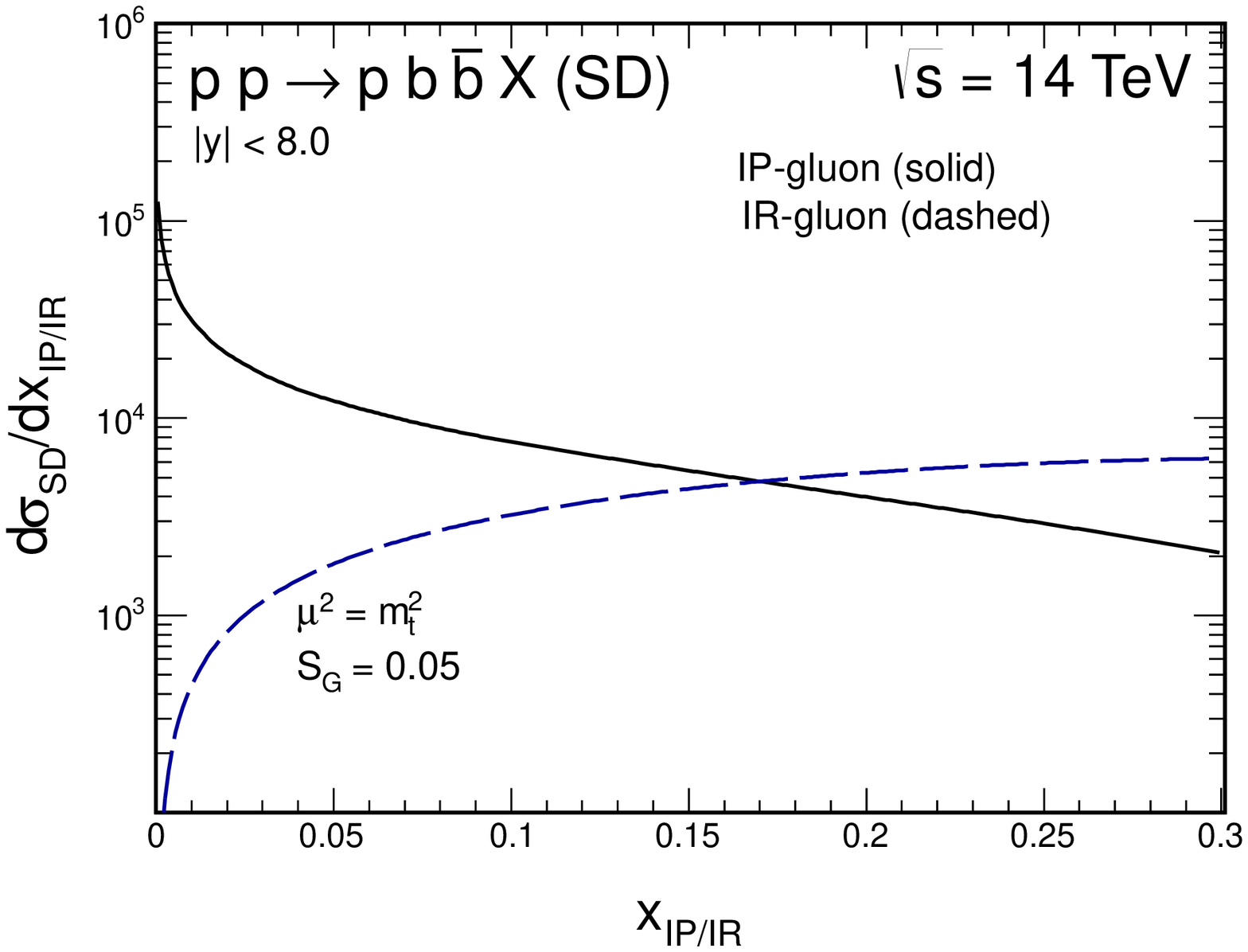}}
\end{minipage}
   \caption{
\small The distribution in $x_\Pom$ (solid) and $x_\Reg$ (dashed) for 
$\sqrt{s}$ = 14 TeV.
The left panel shows distribution in pomeron/reggeon longitudinal momentum fraction for $c$ quarks (antiquarks), the right panel shows similar distributions for $b$ quarks (antiquarks) for single-diffractive production. 
}
 \label{fig:x}
\end{figure}

\begin{figure}[!h]
\begin{minipage}{0.47\textwidth}
 \centerline{\includegraphics[width=1.0\textwidth]{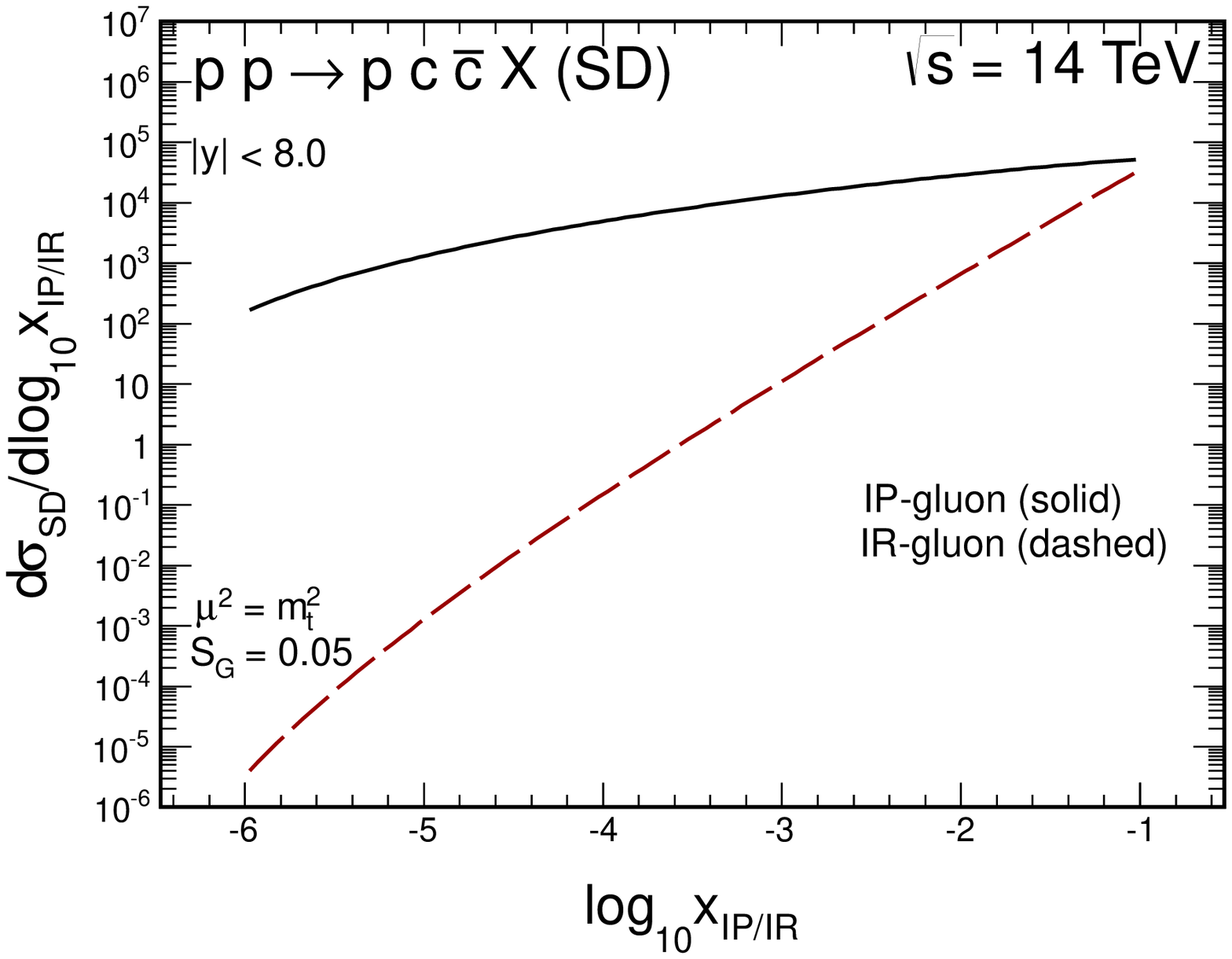}}
\end{minipage}
\hspace{0.5cm}
\begin{minipage}{0.47\textwidth}
 \centerline{\includegraphics[width=1.0\textwidth]{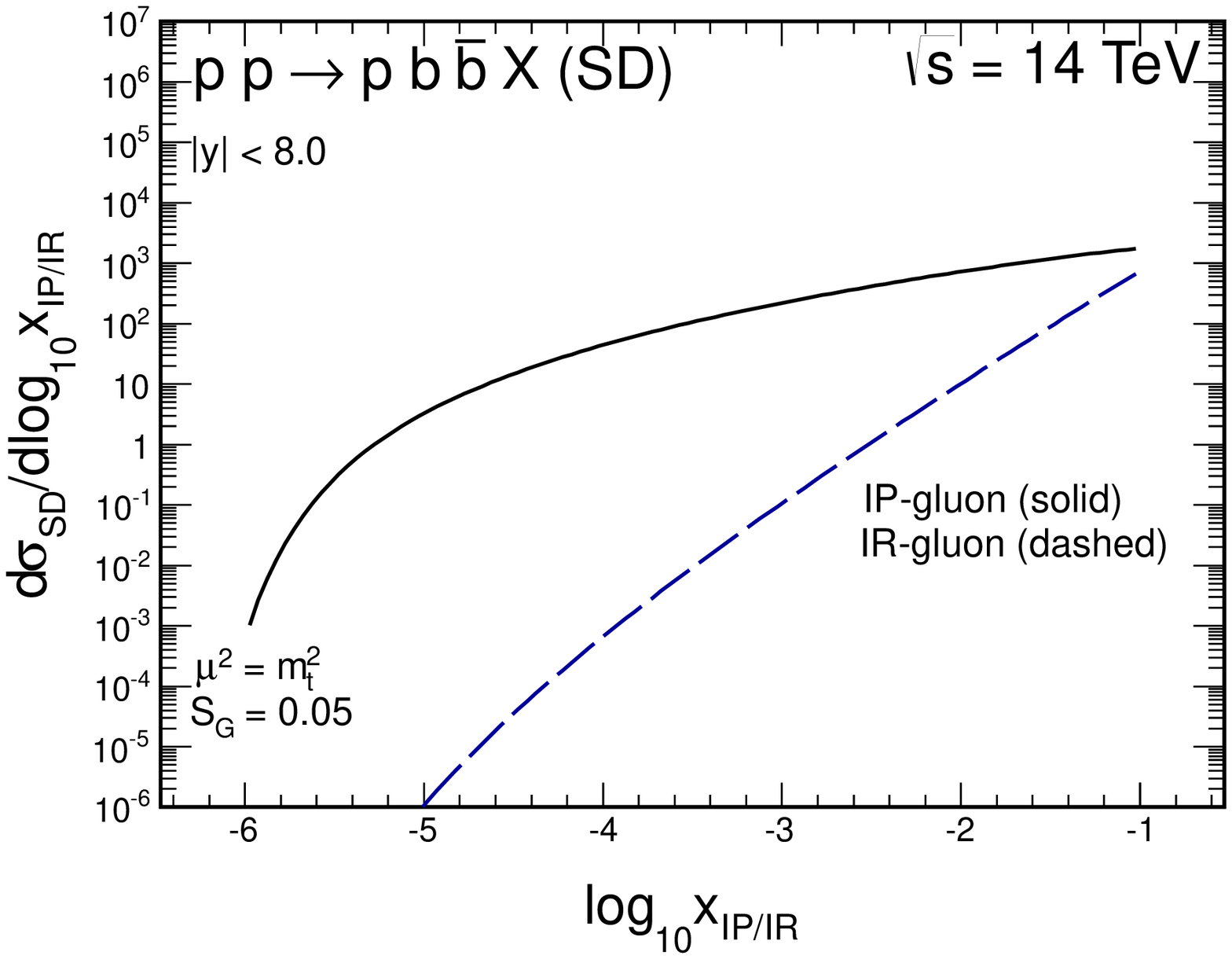}}
\end{minipage}
   \caption{
\small The distribution in $\log_{10}x_{\Pom}$ (solid) and $\log_{10}x_{\Reg}$ (dashed) for 
$\sqrt{s}$ = 14 TeV.
The left panel shows distribution for $c$ quarks (antiquarks)
and the right panel for $b$ quarks (antiquarks) distribution for single-diffractive production.
}
 \label{fig:xlog}
\end{figure}

In Fig.~\ref{fig:pt_x} we show separately contributions for different upper limits for the value of $x_\Pom$ and $x_\Reg$. The shape of these distributions are rather similar.
As a default, in the case of pomeron exchange the upper limit in 
the convolution formula is taken to be 0.1 and for reggeon exchange 0.2.
Additionally Fig.~\ref{fig:x} shows distribution in pomeron/reggeon longitudinal momentum fraction for $c$ quarks (antiquarks) (left panel) and for $b$ quarks (antiquarks) (right panel) for single-diffractive production. The similar distributions in $\log_{10}x_{\Pom}$ and $\log_{10}x_{\Reg}$  are presented in Fig.~\ref{fig:xlog}. 
In our opinion, the whole Regge formalism does not apply above these limits
and therefore unphysical results could be obtained.
%

\begin{figure}[!h]
\begin{minipage}{0.47\textwidth}
 \centerline{\includegraphics[width=1.0\textwidth]{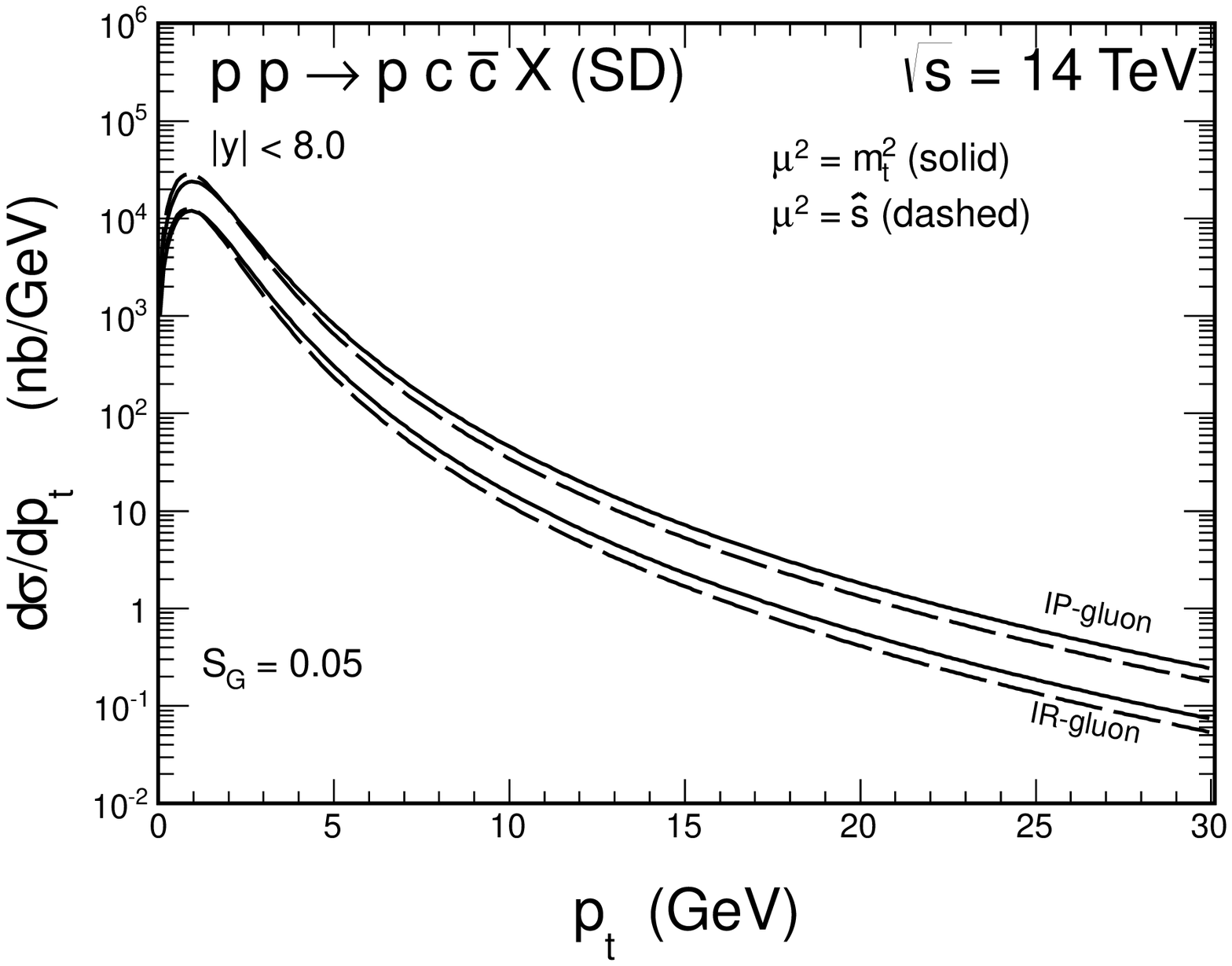}}
\end{minipage}
\hspace{0.5cm}
\begin{minipage}{0.47\textwidth}
 \centerline{\includegraphics[width=1.0\textwidth]{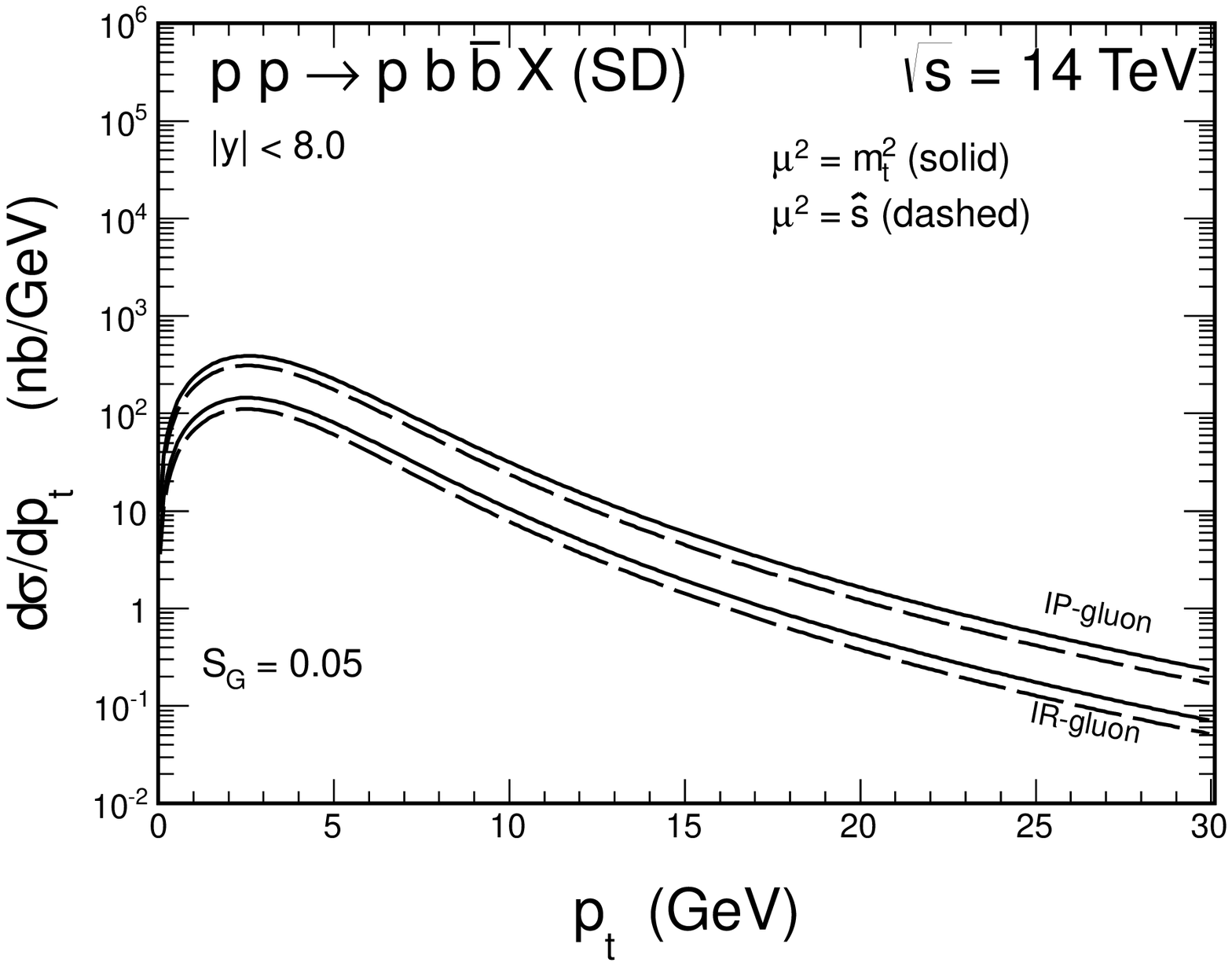}}
\end{minipage}
   \caption{
\small Transverse momentum distribution of $c$ quarks (antiquarks) (left)
and $b$ quarks (antiquarks) (right) for single-diffractive production at $\sqrt{s} = 14$ TeV
for factorization scales: $\mu^2 = m_t^2$ (solid) and $\mu^2 = \hat{s}$ (dashed). 
}
 \label{fig:pt_scale}
\end{figure}

For completeness, in Fig.~\ref{fig:pt_scale} 
we show separately contributions for different factorization scales: $\mu^2 = m_t^2$ and $\mu^2 = \hat{s}$,
which give quite similar distributions in
transverse momentum.

\begin{figure}[!h]
\begin{minipage}{0.47\textwidth}
 \centerline{\includegraphics[width=1.0\textwidth]{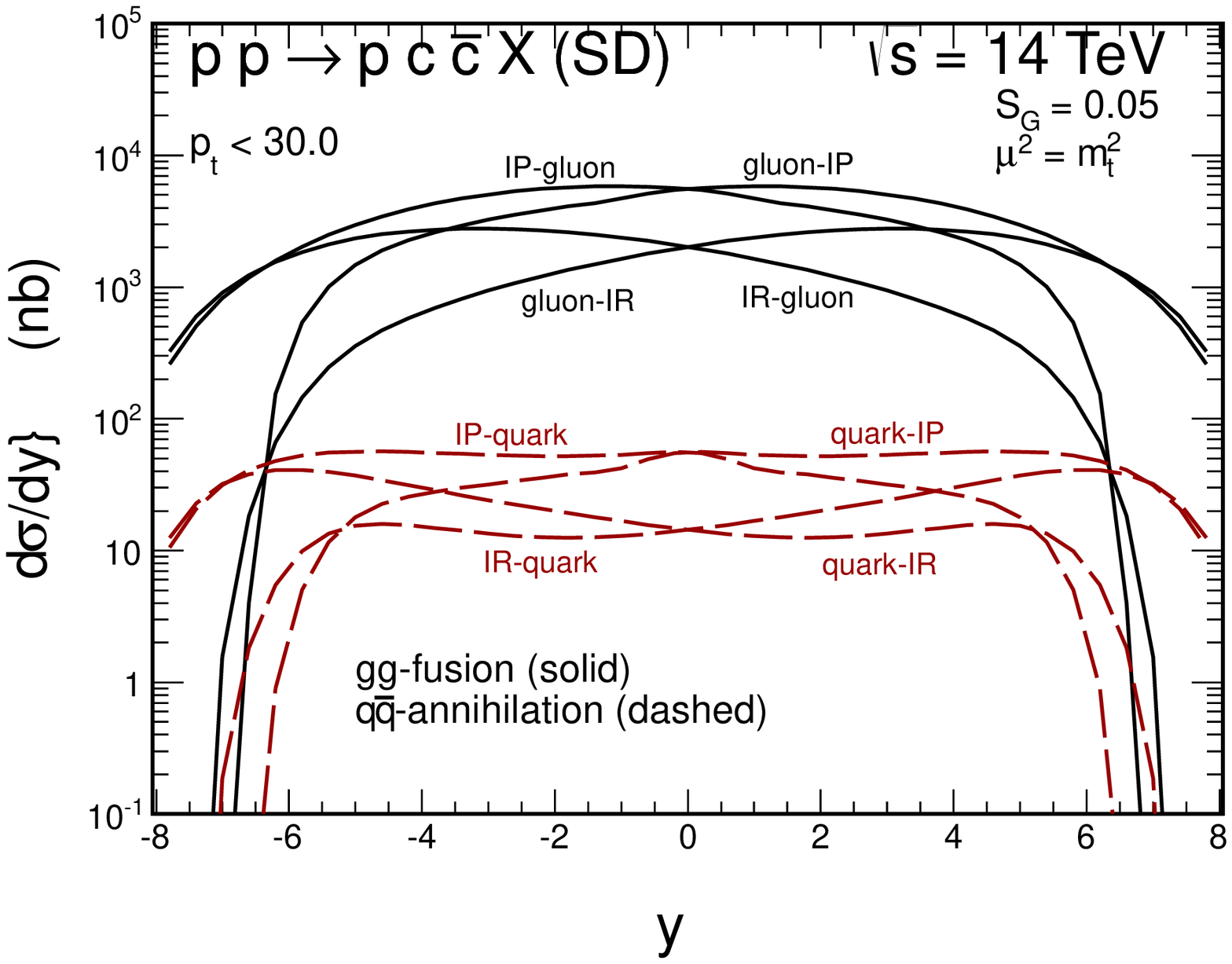}}
\end{minipage}
\hspace{0.5cm}
\begin{minipage}{0.47\textwidth}
 \centerline{\includegraphics[width=1.0\textwidth]{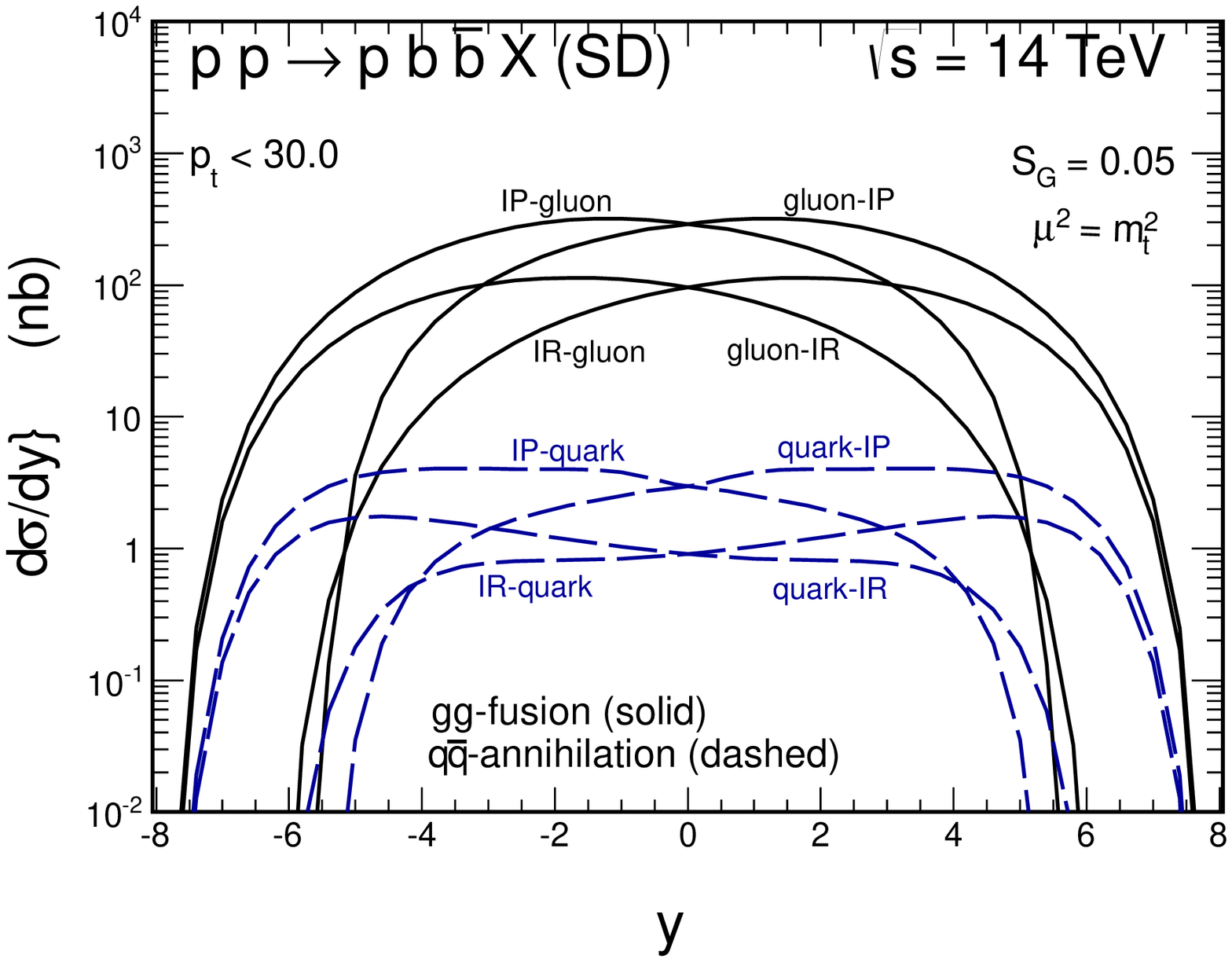}}
\end{minipage}
   \caption{
\small Rapidity distribution of $c$ quarks (antiquarks) (left)
and $b$ quarks (antiquarks) (right) for single-diffractive production at $\sqrt{s} = 14$ TeV.
Components of the pomeron-gluon (and gluon-pomeron), the pomeron-quark(antiquark) (and quark(antiquark)-pomeron) and the reggeon-gluon (and gluon-reggeon),
the reggeon-quark(antiquark) (and quark(antiquark)-reggeon) mechanisms are shown separately.
}
 \label{fig:y}
\end{figure}

\begin{figure}[!h]
\begin{minipage}{0.47\textwidth}
 \centerline{\includegraphics[width=1.0\textwidth]{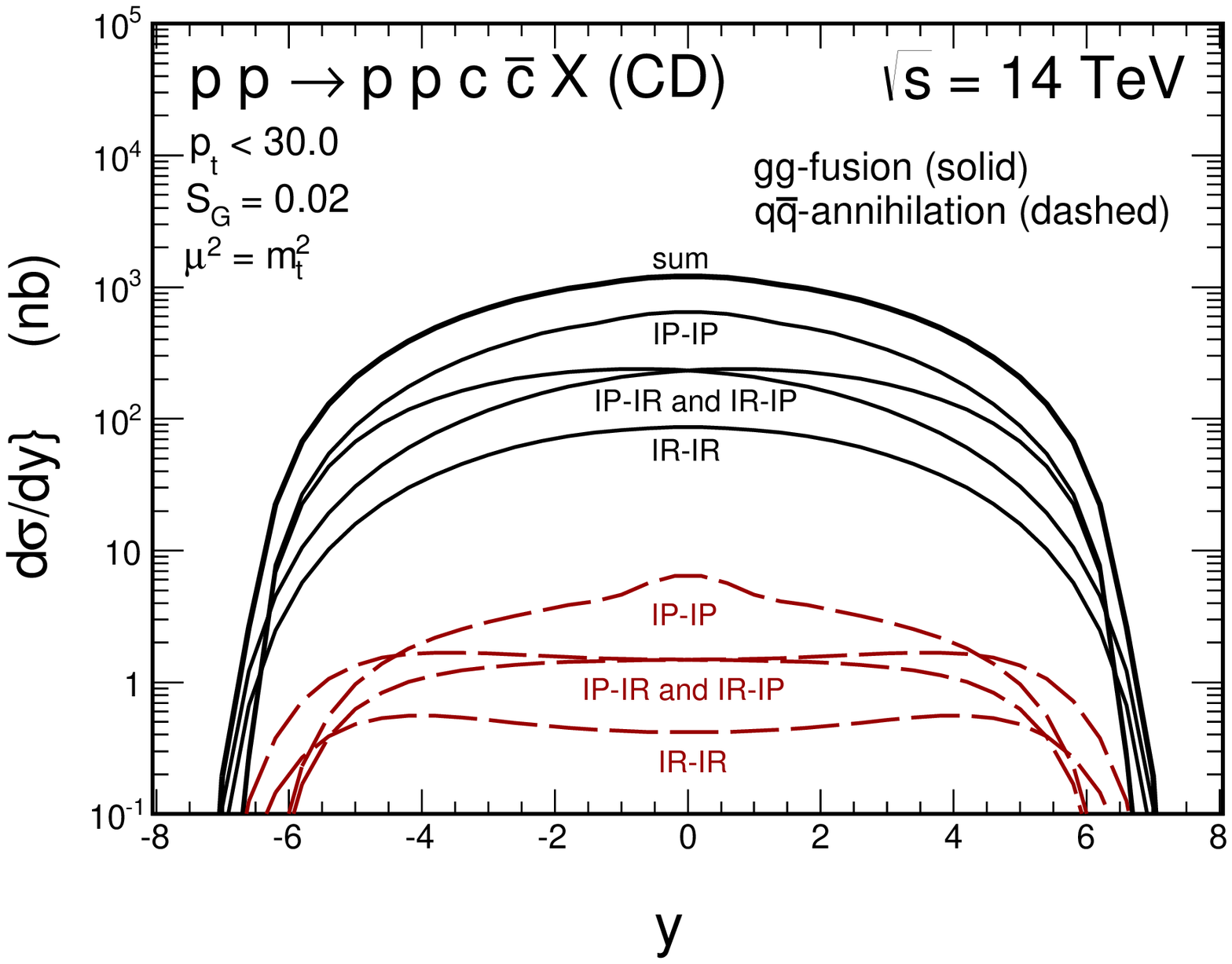}}
\end{minipage}
\hspace{0.5cm}
\begin{minipage}{0.47\textwidth}
 \centerline{\includegraphics[width=1.0\textwidth]{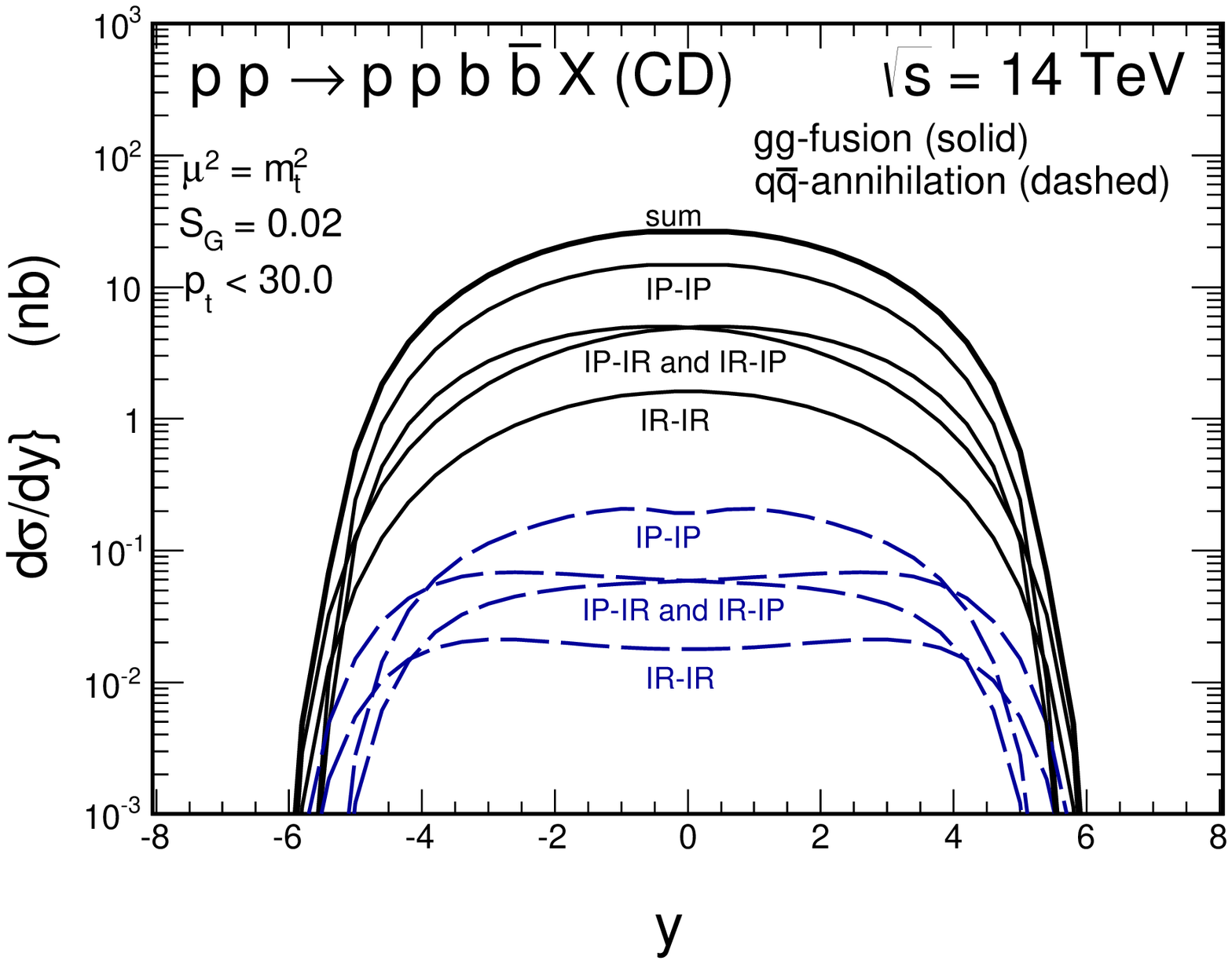}}
\end{minipage}
   \caption{
\small Rapidity distribution of $c$ quarks (antiquarks) (left)
and $b$ quarks (antiquarks) (right) for the central-diffractive production at $\sqrt{s} = 14$ TeV.
Components of the pomeron-pomeron, reggeon-reggeon, pomeron-reggeon and reggeon-pomeron mechanisms are shown separately.
The sum of all contributions is shown by the thick solid line. 
}
 \label{fig:y_CD}
\end{figure}

Figures~\ref{fig:y} and~\ref{fig:y_CD} show rapidity distributions for
$c$ quarks (antiquarks) (left panels)
and $b$ quarks (antiquarks) (right panels) pair production for single- and central-diffractive mechanisms respectively. 
The rapidity distributions for pomeron-gluon (and gluon-pomeron), pomeron-quark(antiquark) (and quark(antiquark)-pomeron) and reggeon-gluon (and gluon-reggeon),
reggeon-quark(antiquark) (and quark(antiquark)-reggeon
mechanisms in the single-diffractive case are shifted
to forward and backward rapidities, respectively. 
The distributions for the individual single-diffractive
mechanisms have maxima at large rapidities, while
the central-diffractive contribution is concentrated
at midrapidities. This is a consequence of limiting integration
over $x_\Pom$ in Eq.(\ref{flux_of_Pom}) to 0.0 $< x_\Pom <$ 0.1 
and over $x_\Reg$ to 0.0 $< x_\Reg <$ 0.2.

\begin{figure}[!h]
\begin{minipage}{0.47\textwidth}
 \centerline{\includegraphics[width=1.0\textwidth]{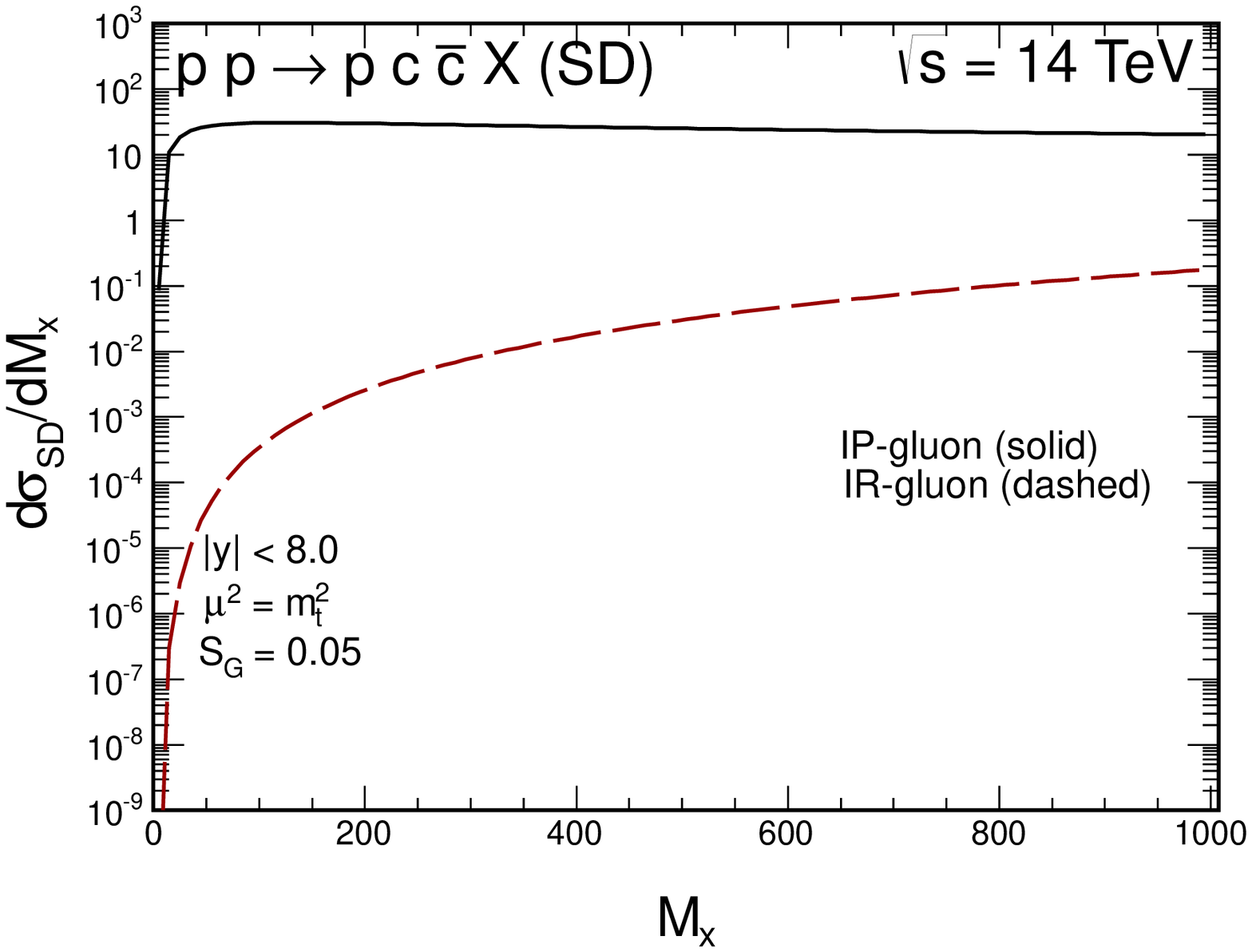}}
\end{minipage}
\hspace{0.5cm}
\begin{minipage}{0.47\textwidth}
 \centerline{\includegraphics[width=1.0\textwidth]{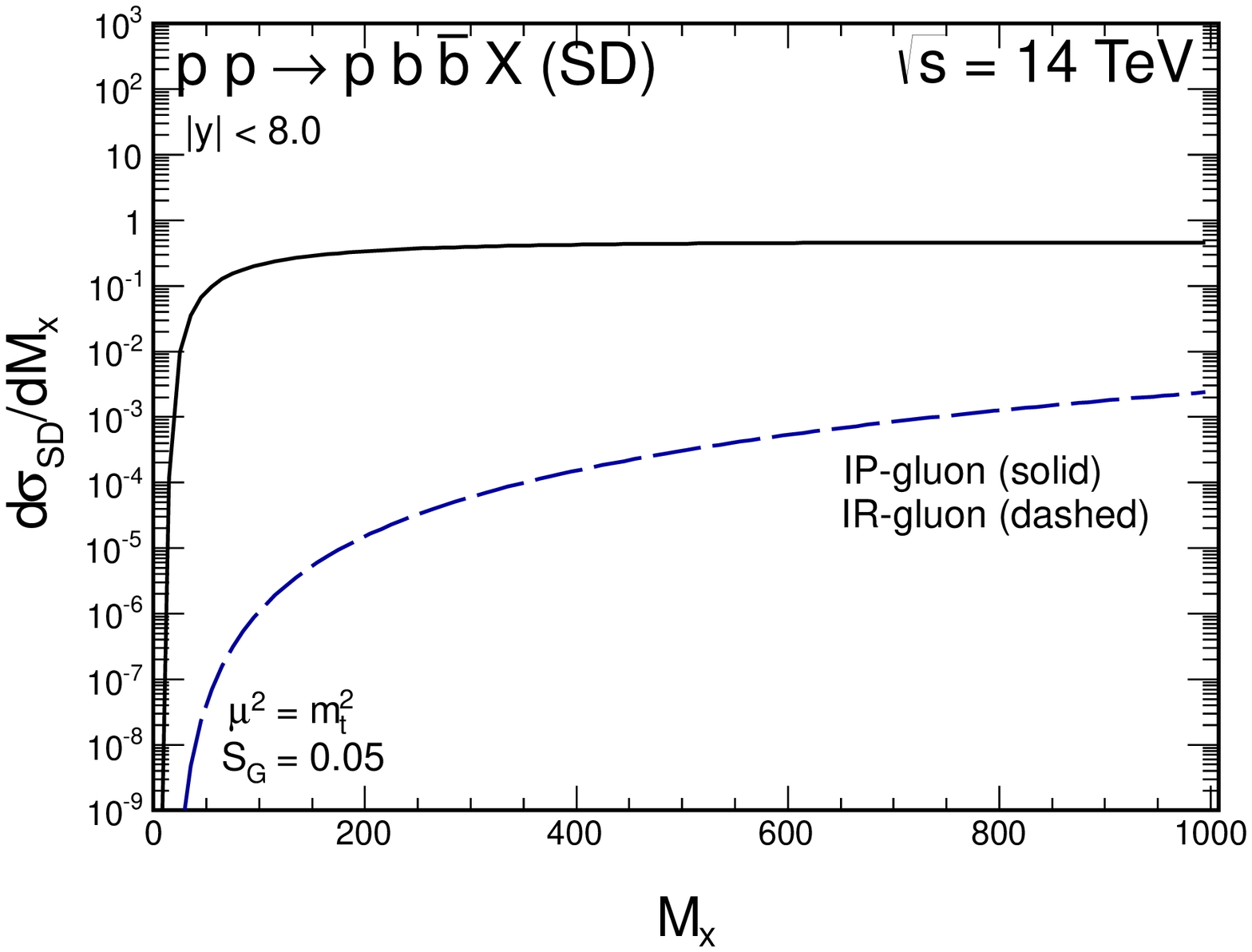}}
\end{minipage}
   \caption{
\small The missing mass distribution for 
$\sqrt{s}$ = 14 TeV.
The left panel shows distribution for $c$ quarks (antiquarks),
the right panel $b$ quarks (antiquarks) production for single-diffractive production.
}
 \label{fig:Mx}
\end{figure}

Finally, In Fig.~\ref{fig:Mx} we show the missing mass distribution for
$c$ quarks (antiquarks) (left panel) and
for $b$ quarks (antiquarks) (right panel) for single-diffractive production.
These both contributions have similar shapes of distributions. 
Experimentally, measuring the distributions
in invariant mass of $D$ and $B$ mesons
would be interesting and will be discussed in the next section.

\subsection{Heavy quark hadronization effects}

The transition from quarks and gluons to hadrons, called hadronization or parton fragmentation, can be so far approached only through phenomenological models. In principle, in the case of many-particle final states the Lund string model \cite{Andersson:1983ia} and the cluster fragmentation model \cite{Webber:1983if} are usually used, providing good description of the hadronization of the parton system as a whole. However, the hadronization of heavy quarks is usually done with the help of fragmentation functions (FFs) extracted from $e^+ e^-$ experiments (see e.g. Refs.~\cite{Cacciari:2005rk,Cacciari:2012ny,Maciula:2013wg}).

Especially in the case of diffractive production, where one or both protons remain intact, the applicability of the compound hadronization models (implemented in Monte Carlo generators and dedicated to non-diffractive processes) is still an open question. More detailed studies, e.g. of gluonic and quark jet structures in diffractive events, are needed to draw more definite conclusions in this context. In our calculation we follow the fragmentation function technique which seems to be sufficient to make first evaluation of corresponding cross sections. This scheme has been recently successfully used for description of inclusive non-diffractive open charm and bottom data at the LHC \cite{Cacciari:2012ny,Maciula:2013wg}. In the context of diffractive production studies, the uncertainties coming from the process of parton fragmentation seem to be less important than those related to the parton-level diffractive calculation (e.g. uncertainties of diffractive PDFs or gap survival probability).

\begin{figure}[!h]
\begin{minipage}{0.47\textwidth}
 \centerline{\includegraphics[width=1.0\textwidth]{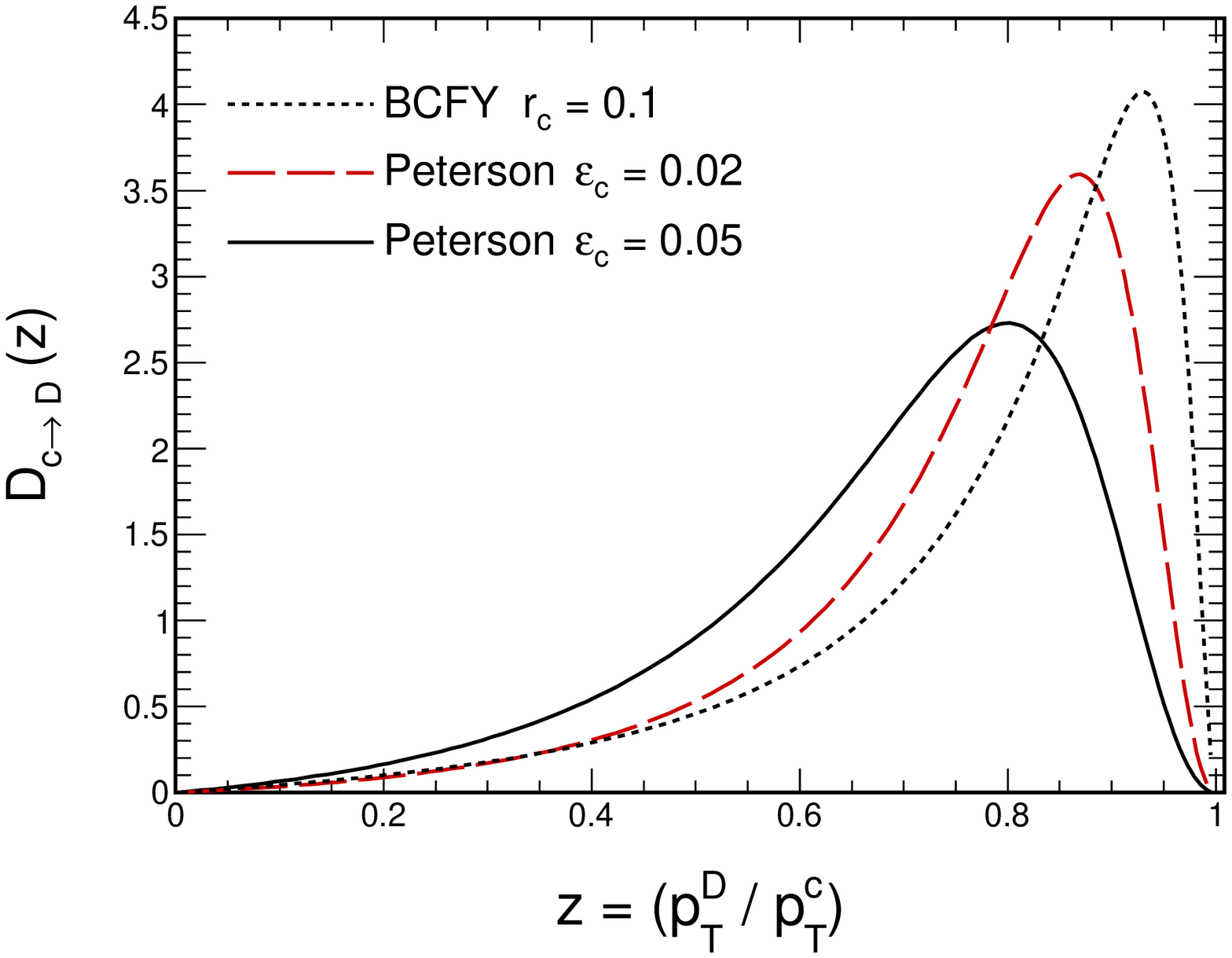}}
\end{minipage}
\hspace{0.5cm}
\begin{minipage}{0.47\textwidth}
 \centerline{\includegraphics[width=1.0\textwidth]{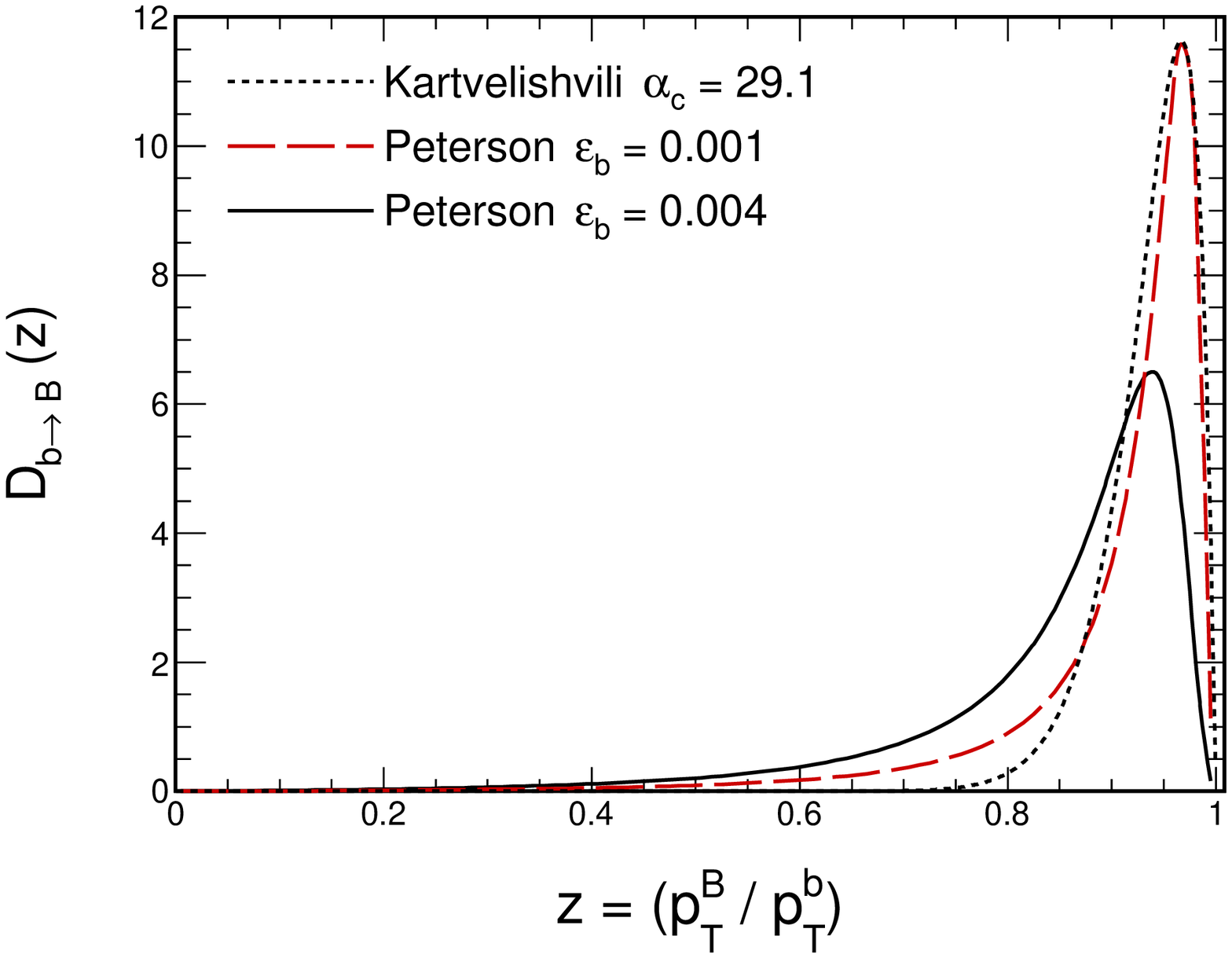}}
\end{minipage}
   \caption{
\small Different models of the fragmentation functions for charm (left) and bottom (right) quarks. The default functions from the FONLL framework are compared to the Peterson functions with different $\varepsilon$ parameters.}
 \label{fig:z-FF}
\end{figure}

According to the fragmentation function formalism, in the following numerical calculations,
the differential distributions of open charm and bottom hadrons $h =D, B$, e.g. for single-diffractive production, are obtained through a convolution of differential distributions of heavy quarks/antiquarks and $Q \to h$ fragmentation functions:
\begin{equation}
\frac{d \sigma(pp \rightarrow h \bar{h} \; p X)}{d y_h d^2 p_{t,h}} \approx
\int_0^1 \frac{dz}{z^2} D_{Q \to h}(z)
\frac{d \sigma(pp \rightarrow Q \overline{Q} \; p X)}{d y_Q d^2 p_{t,Q}}
\Bigg\vert_{y_Q = y_h \atop p_{t,Q} = p_{t,h}/z} \;,
\label{Q_to_h}
\end{equation}
where $p_{t,Q} = \frac{p_{t,h}}{z}$ and $z$ is the fraction of
longitudinal momentum of heavy quark $Q$ carried by a hadron $h$.
Technically, in this scheme of fragmentation the rescalling of the transverse momentum is the most important effect. This is because one needs to deal with very steep functions of transverse momenta. Since the rapidity spectra are usually flat, or slowly varying, the approximation assuming that $y_{Q}$ is
unchanged in the fragmentation process, i.e. $y_h = y_Q$, is commonly applied. This approximation is typical for light hadrons, however, is also commonly accepted for heavy quarks, especially in the region of not too small quark $p_{t}$'s. The fragmentation functions for heavy quarks are peaked at large $z$ (see Fig.~\ref{fig:z-FF}) so the problematic small-$p_{t}$ region is suppressed. 

In all the following numerical calculations the standard Peterson fragmentation function \cite{Peterson:1982ak} is applied. The default set of the parameters for these functions is $\varepsilon_{c} = 0.05$ for charm and $\varepsilon_{b} = 0.004$ for bottom quarks, respectively. This values were extracted by H1 \cite{Aaron:2008ac}, ALEPH \cite{Heister:2001jg} and OPAL \cite{Abbiendi:2002vt} analyses. However, in the similar fragmentation scheme applied in the FONLL framework for hadroproduction of heavy flavours at RHIC \cite{Cacciari:2005rk} and LHC \cite{Cacciari:2012ny}, rather harder functions are suggested. Within the FONLL approach the Braaten-Cheung-Fleming-Yuan (BCFY) \cite{Braaten:1994bz} function with $r_c = 0.1$ for charm and the Kartvelishvili \cite{Kartvelishvili:1977pi} parametrization with $\alpha_b = 29.1$ for bottom are used. In our calculation, to make the shapes of the Peterson functions 
closer to those from the FONLL approach, the parameters are fixed to $\varepsilon_{c} = 0.02$ and $\varepsilon_{b} = 0.001$ (see Fig.~\ref{fig:z-FF}). In the following numerical predictions of the cross sections for $D^{0}$ and $B^{\pm}$ mesons the fragmentation functions are normalized to the branching fractions
from Refs.~\cite{Lohrmann:2011np,Beringer:1900zz,Aaltonen:2008zd}, i.e. $\textrm{BR}(c \to D^{0}) = 0.565$ and $\textrm{BR}(b \to B^{\pm}) = 0.4$.

\subsection{Cross sections for $D^{0}$ and $B^{\pm}$ mesons production}

Measurements of charm and bottom cross sections at hadron colliders can be performed in the so-called direct way.
This method is based on full reconstruction of all decay products of open charm and bottom mesons, for instance in the 
$D^0 \to K^- \pi^+ $, $D^+ \to K^- \pi^+ \pi^+$ or $B^+ \to J/\psi K^+ \to K^+ \mu^+ \mu^-$ channels. The decay products
with an invariant mass from the expected hadron decay combinations, permit direct observation of $D$ or $B$ meson as a peak in invariant mass spectrum. Then, after a substraction of invariant mass continuum the relevant cross section for the meson production can be provided.
The same method can be applied for measurement of charm and bottom production rates for the diffractive events.

\begin{table}[tb]%
\caption{Integrated cross sections for diffractive production of open charm and bottom mesons in different measurement modes for ATLAS, LHCb and CMS experiments at $\sqrt{s}=14$ TeV.}
\newcolumntype{Z}{>{\centering\arraybackslash}X}
\label{table}
\centering %
\resizebox{\textwidth}{!}{%
\begin{tabularx}{16.cm}{ZZZZZ}
\toprule[0.1em] %
\\[-2.4ex] 
\\[-7.0ex]

\multirow{3}*{Acceptance} & \multirow{3}*{Mode} & \multicolumn{3}{c}{\underline{$\;\;\;\;\;\;\;\;\;\;\;\;\;\;\;\;\;\;\;\;$ Integrated cross sections, [nb] $\;\;\;\;\;\;\;\;\;\;\;\;\;\;\;\;\;\;\;\;$ }}  \\[+0.8ex]
                          &                           &  \multirow{2}*{single-diffractive}  & \multirow{2}*{central-diffractive}  & non-diffractive\\[-2.ex]
                          &                           &                                     &                                     & EXP data\\[-2.ex]
                          
\toprule[0.1em]\\[-4.4ex]
ATLAS, $|y|< 2.5$           &  \multirow{2}*{$D^{0}+\overline{D^{0}}$} &  \multirow{2}*{3555.22 ($I\!R\!: 25\%$)} & \multirow{2}*{177.35 ($I\!R\!: 43\%$)} & \multirow{2}*{$-$}\\[+0.4ex]
$p_{\perp} > 3.5$ GeV                          & & & & \\[-1.ex] 
LHCb,$\;$~$2~\!<~\!y~\!<~\!4.5$           &  \multirow{2}*{$D^{0}+\overline{D^{0}}$} &  \multirow{2}*{31442.8 ($I\!R\!: 31\%$)} & \multirow{2}*{2526.7 ($I\!R\!: 50\%$)} & \multirow{2}*{$1488000\pm182000$}\\[+0.4ex]
$p_{\perp} < 8$ GeV                          & & & & \\[-1.ex] 
\hline \\[-4.4ex]
CMS, $|y|< 2.4$           &  \multirow{2}*{$(B^{+}+B^{-})/2$} &  \multirow{2}*{349.18 ($I\!R\!: 24\%$)} & \multirow{2}*{14.24 ($I\!R\!: 42\%$)} & \multirow{2}*{$28100\pm2400\pm2000$}\\[+0.4ex]
$p_{\perp} > 5$ GeV                          & & & & \\[-1.ex] 
LHCb,$\;$~$2~\!<~\!y~\!<~\!4.5$           &  \multirow{2}*{$B^{+}+B^{-}$} &  \multirow{2}*{867.62 ($I\!R\!: 27\%$)} & \multirow{2}*{31.03 ($I\!R\!: 43\%$)} & \multirow{2}*{$41400\pm1500\pm3100$}\\[+0.4ex]
$p_{\perp} < 40$ GeV                          & & & & \\[-1.ex]  

\hline \\[-4.4ex]
LHCb, $2 < y < 4$           &  \multirow{2}*{$D^{0}\overline{D^{0}}$} &  \multirow{2}*{179.4 ($I\!R\!: 28\%$)} & \multirow{2}*{7.67 ($I\!R\!: 45\%$)} & \multirow{2}*{$6230\pm120\pm230$}\\[+0.4ex]
$3 < p_{\perp} < 12$ GeV                          & & & & \\[-1.ex]

\bottomrule[0.1em]
 
\end{tabularx}
}
\end{table}

Numerical predictions of the integrated cross sections for the single- and central-diffractive production of $D^{0}$ and $B^{\pm}$ mesons, including relevant experimental acceptance of the ATLAS, LHCb and CMS detectors,  are collected in Table~\ref{table}. The kinematical cuts are taken to be identical to those which have been already used in the standard non-diffractive measurements of open charm and bottom production rates at the LHC. The corresponding experimental cross sections for non-diffractive processes are shown for reference. In the case of inclusive production of single $D$ or $B$ meson
the ratio of the diffractive integrated cross sections to the non-diffractive one is about $\sim 2\%$ for single- and only about $\sim 0.07\%$ for central-diffractive mechanism. This ratio is only slightly bigger for $D^{0}\overline{D^{0}}$ pair production, becoming of about $\sim 3\%$ and $0.1\%$, respectively. In addition, the relative contribution of the reggeon-exchange mechanisms to the overall diffractive production cross sections is also shown. This relative contribution is about $\sim 24-31\%$ for single-diffractive ($\frac{I\!R}{I\!P+I\!R}$) and $\sim 42-50\%$ for 
central-diffractive processes ($\frac{I\!PI\!R+I\!RI\!P+I\!RI\!R}{I\!PI\!P+I\!PI\!R+I\!RI\!P+I\!RI\!R}$) for both, charm and bottom flavoured mesons. The ratio does not really change for different measurement modes and different experimental acceptance.        

\begin{figure}[!h]
\begin{minipage}{0.47\textwidth}
 \centerline{\includegraphics[width=1.0\textwidth]{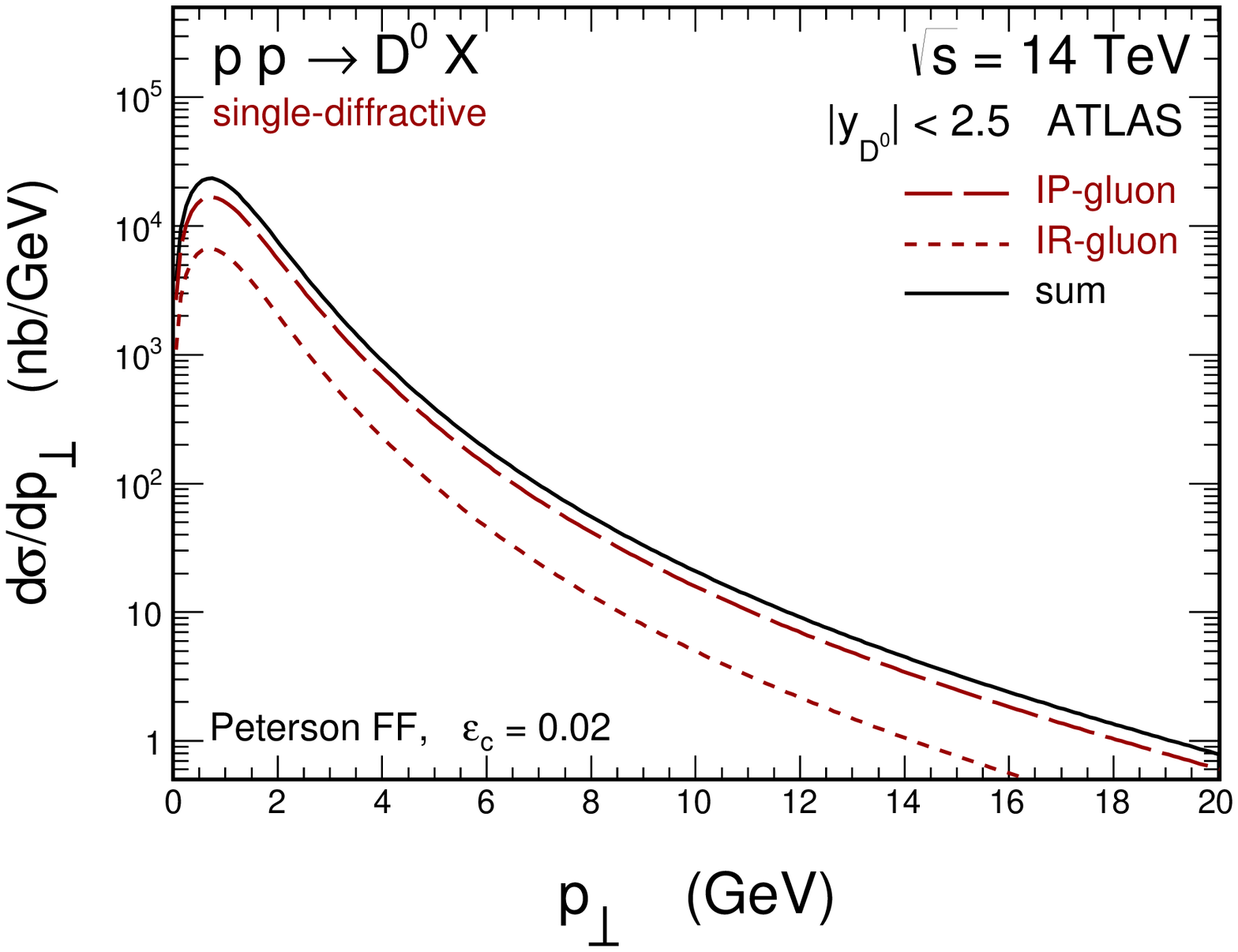}}
\end{minipage}
\hspace{0.5cm}
\begin{minipage}{0.47\textwidth}
 \centerline{\includegraphics[width=1.0\textwidth]{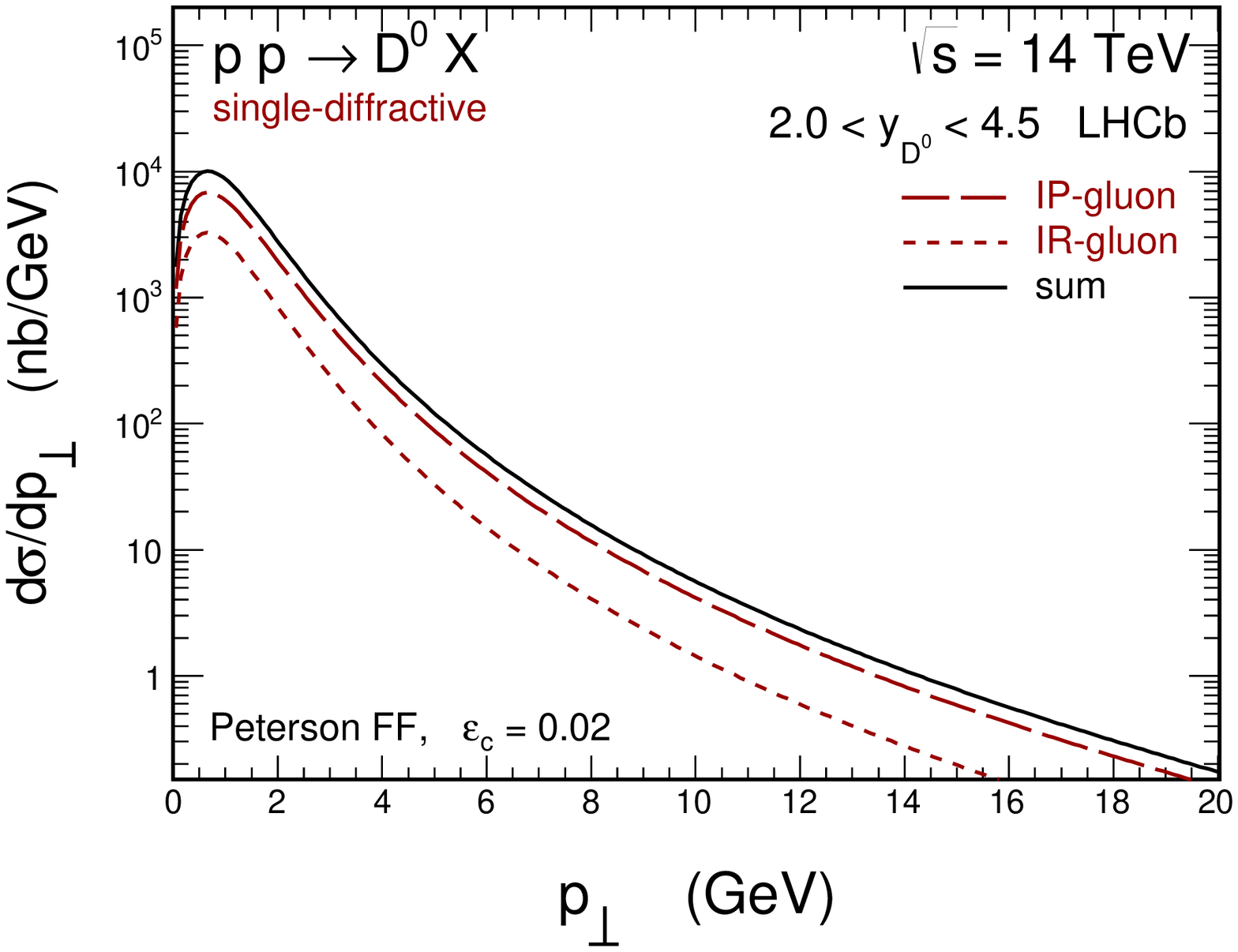}}
\end{minipage}
   \caption{
\small Transverse momentum distribution of $D^{0}$ meson within the ATLAS (left) and the LHCb (right) acceptance for single-diffractive production at $\sqrt{s} = 14$ TeV. Components of the pomeron-gluon (and gluon-pomeron) (long-dashed line) and the reggeon-gluon (and gluon-reggeon) (short-dashed line) contributions are shown separately.
}
 \label{fig:pt-SD-Dmes}
\end{figure}

\begin{figure}[!h]
\begin{minipage}{0.47\textwidth}
 \centerline{\includegraphics[width=1.0\textwidth]{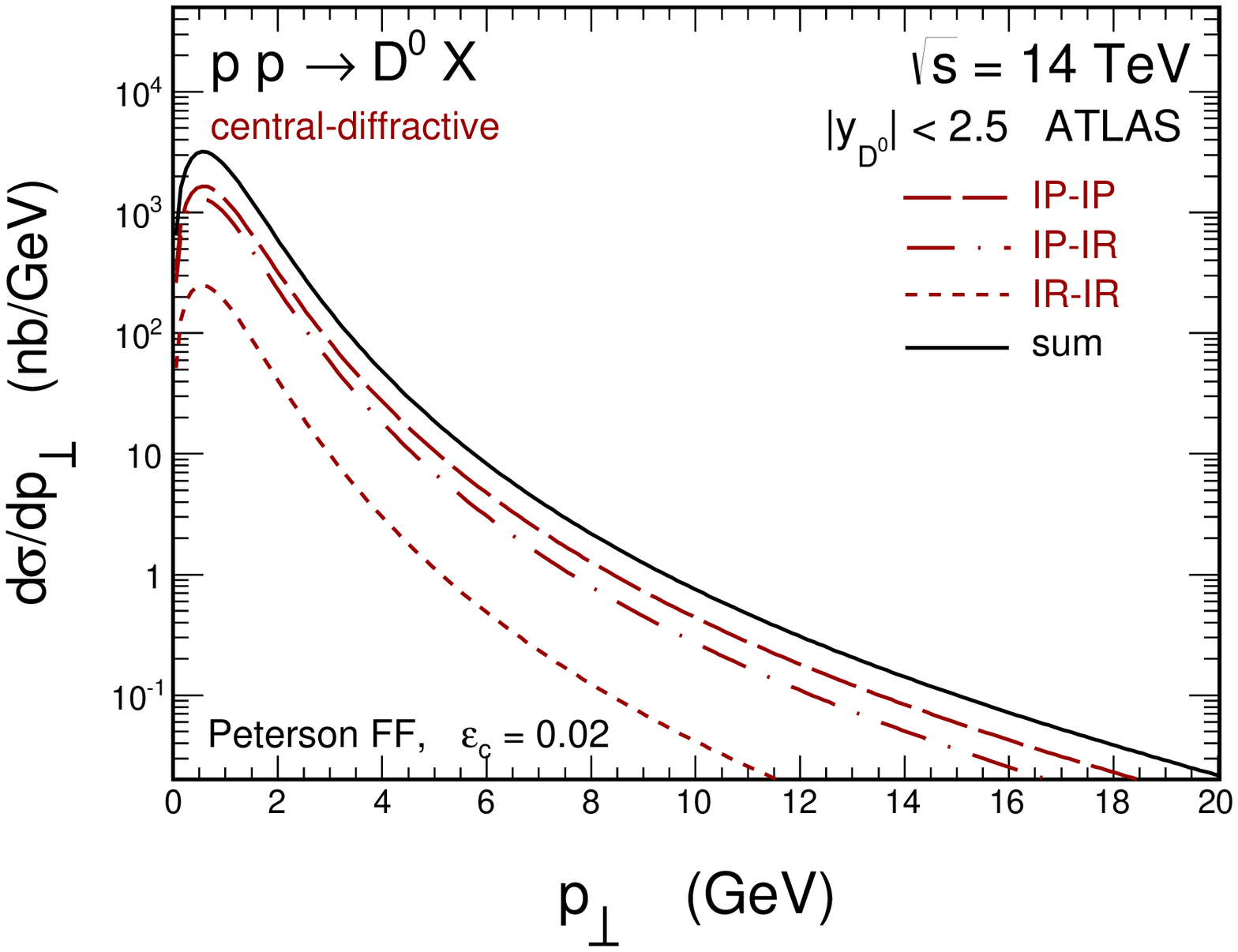}}
\end{minipage}
\hspace{0.5cm}
\begin{minipage}{0.47\textwidth}
 \centerline{\includegraphics[width=1.0\textwidth]{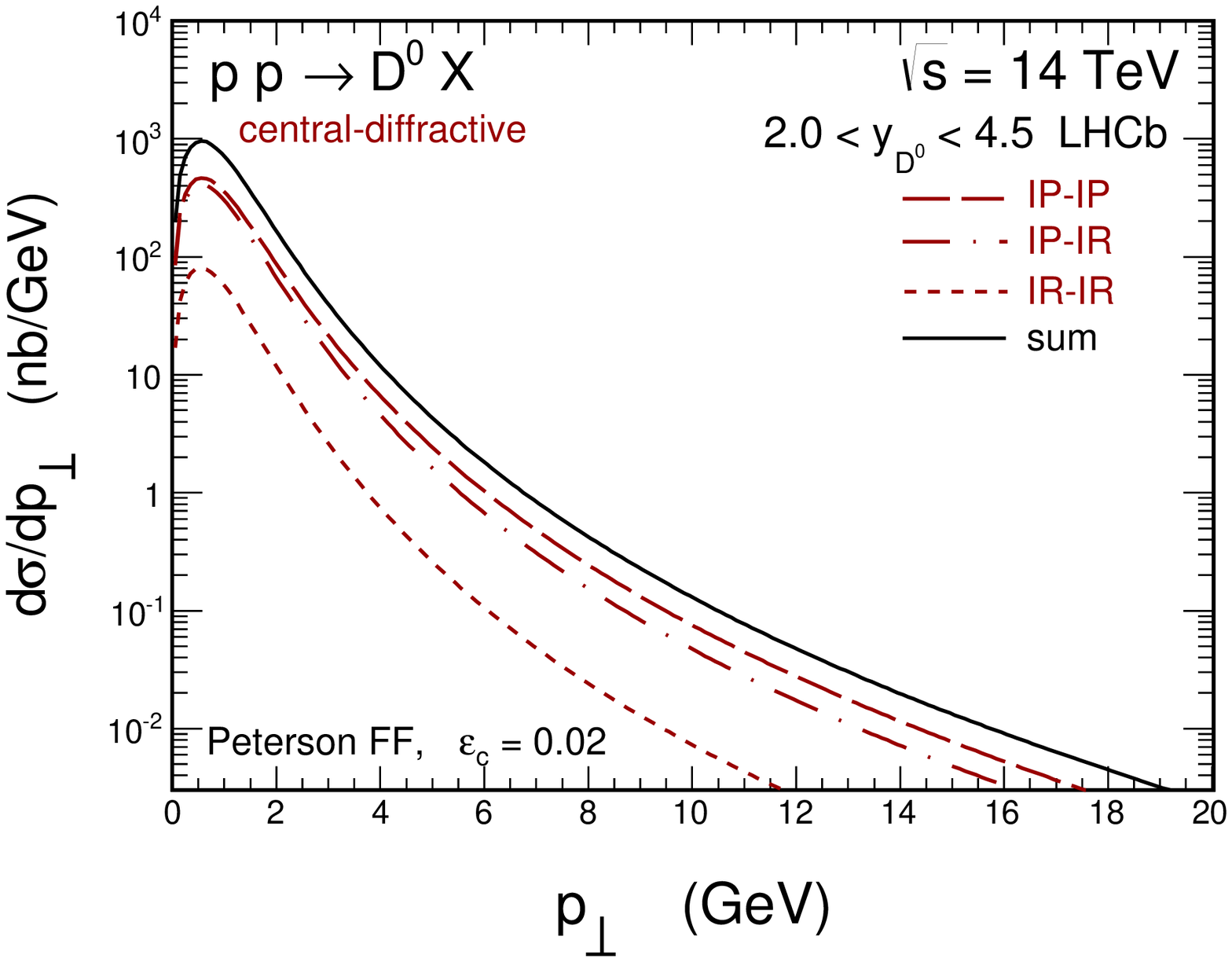}}
\end{minipage}
   \caption{
\small Transverse momentum distribution of $D^{0}$ meson within the ATLAS (left) and the LHCb (right) acceptance for central-diffractive production at $\sqrt{s} = 14$ TeV. Components of the pomeron-pomeron, pomeron-reggeon, reggeon-pomeron and the reggeon-reggeon contributions are shown separately.
}
 \label{fig:pt-CD-Dmes}
\end{figure}

\begin{figure}[!h]
\begin{minipage}{0.47\textwidth}
 \centerline{\includegraphics[width=1.0\textwidth]{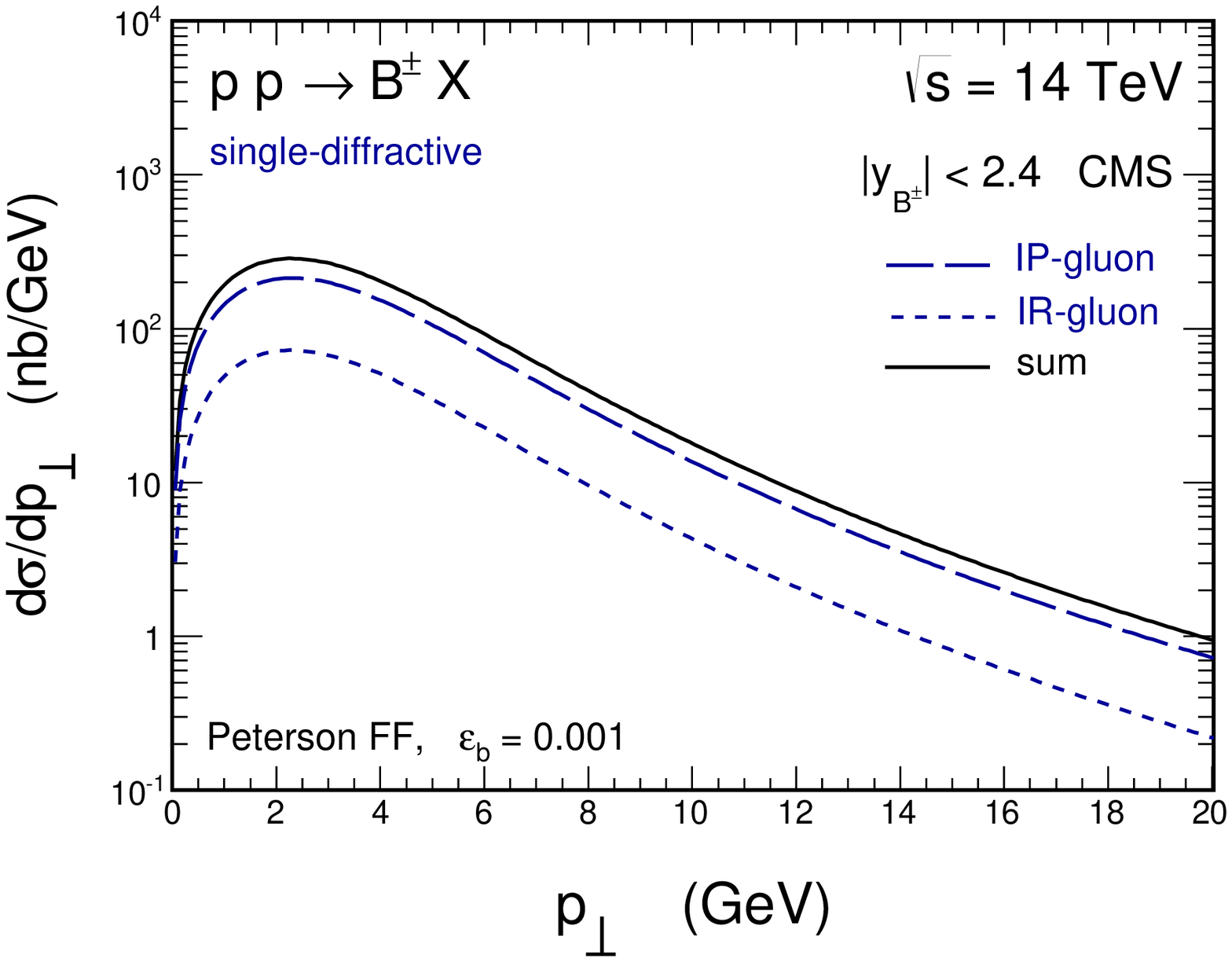}}
\end{minipage}
\hspace{0.5cm}
\begin{minipage}{0.47\textwidth}
 \centerline{\includegraphics[width=1.0\textwidth]{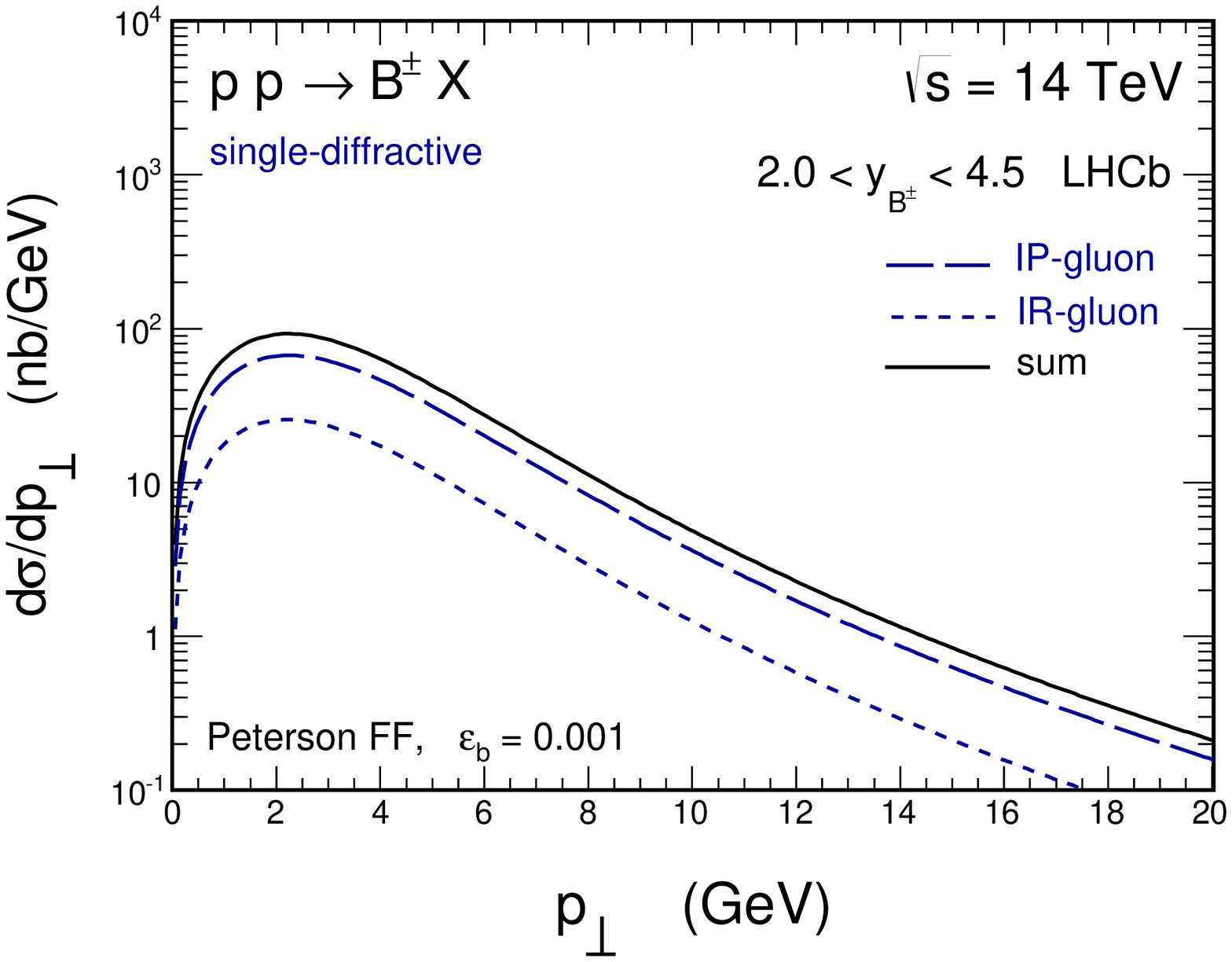}}
\end{minipage}
   \caption{
\small Transverse momentum distribution of $B^{\pm}$ meson within the CMS (left) and the LHCb (right) acceptance for single-diffractive production at $\sqrt{s} = 14$ TeV. Components of the pomeron-gluon (and gluon-pomeron) (long-dashed line) and the reggeon-gluon (and gluon-reggeon) (short-dashed line) contributions are shown separately.
}
 \label{fig:pt-SD-Bmes}
\end{figure}

\begin{figure}[!h]
\begin{minipage}{0.47\textwidth}
 \centerline{\includegraphics[width=1.0\textwidth]{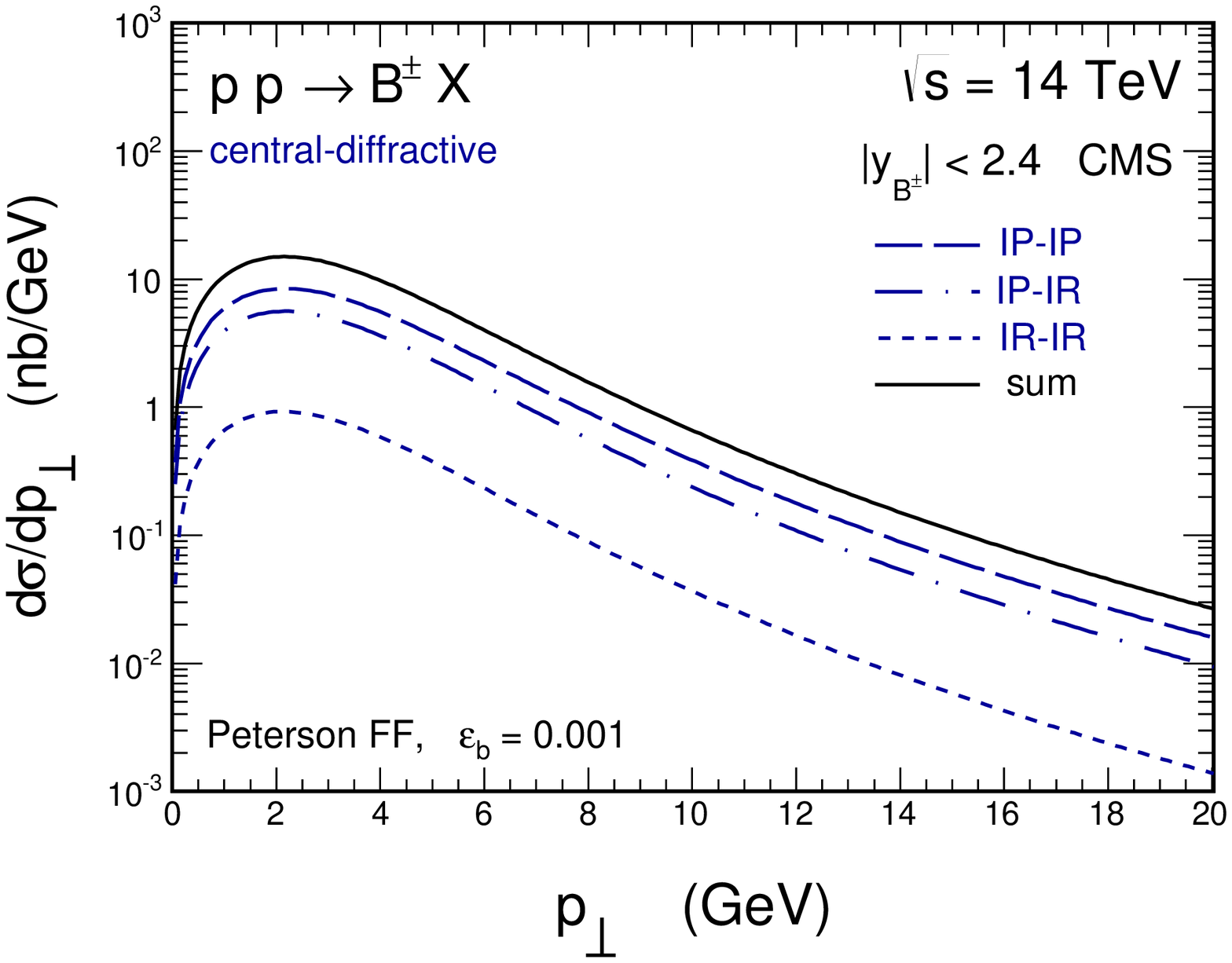}}
\end{minipage}
\hspace{0.5cm}
\begin{minipage}{0.47\textwidth}
 \centerline{\includegraphics[width=1.0\textwidth]{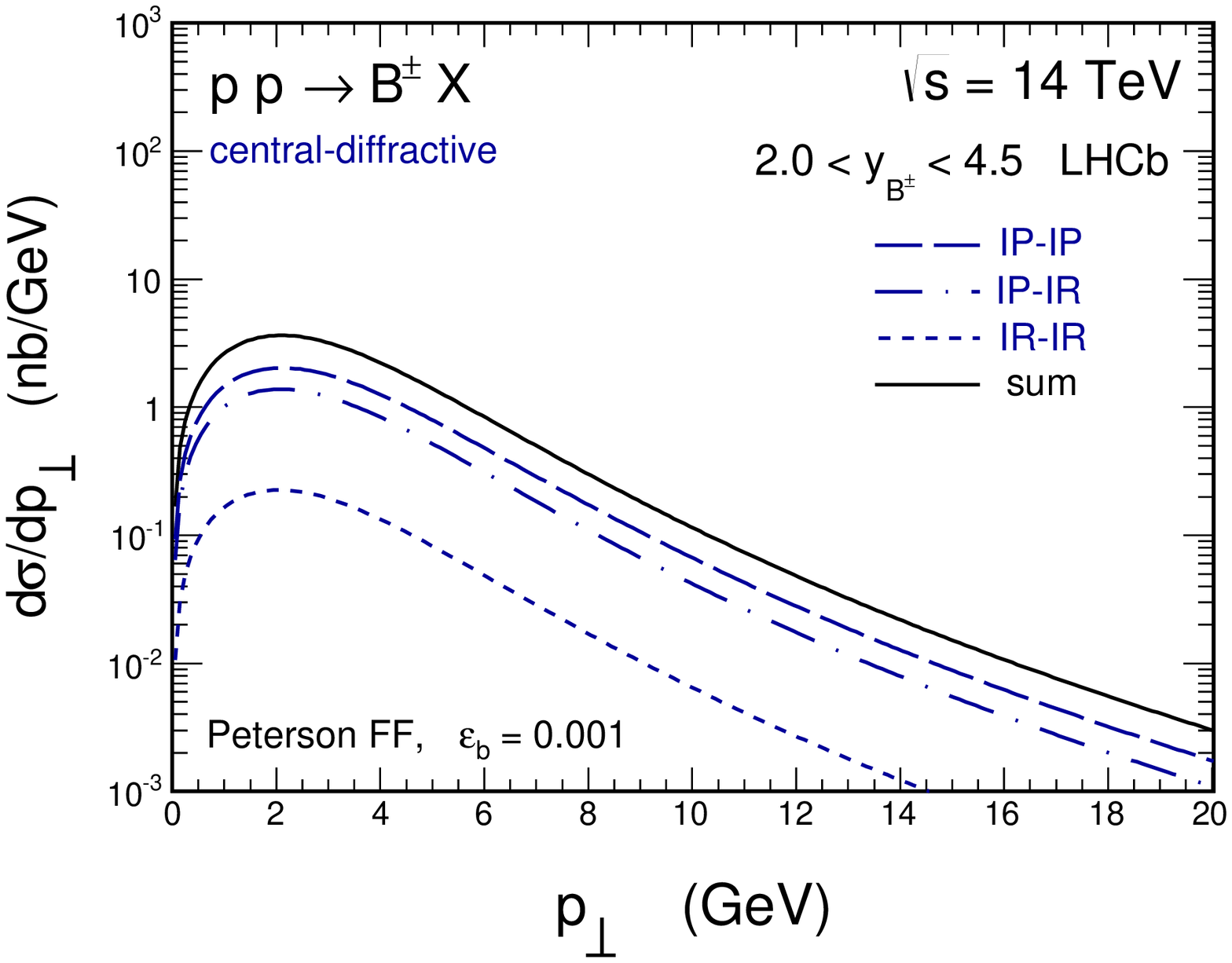}}
\end{minipage}
   \caption{
\small Transverse momentum distribution of $B^{\pm}$ meson within the CMS (left) and the LHCb (right) acceptance for central-diffractive production at $\sqrt{s} = 14$ TeV. Components of the pomeron-pomeron, pomeron-reggeon, reggeon-pomeron and reggeon-reggeon mechanisms are shown separately.
}
 \label{fig:pt-CD-Bmes}
\end{figure}

Figures~\ref{fig:pt-SD-Dmes} and~\ref{fig:pt-CD-Dmes} show transverse momentum distributions of $D^{0}$ meson at $\sqrt{s}=14$ TeV within the ATLAS (left panels) and the LHCb (right panels) acceptance for single- and central-diffractive production, respectively. The contributions of the pomeron- (long-dashed lines) and reggeon-exchange (short-dashed lines) mechanisms are shown separately. These both contributions have similar shapes of the distributions and differ only in normalization. Therefore,  one should not expect a possibility to extract and to test the reggeon component within the special cuts in transverse momentum. The similar distributions (with identical conclusions) but for $B^{\pm}$ meson within the CMS (left panels) and the LHCb (right panels) acceptance are presented in Figs.~\ref{fig:pt-SD-Bmes} and~\ref{fig:pt-CD-Bmes}.   

\begin{figure}[!h]
\begin{minipage}{0.47\textwidth}
 \centerline{\includegraphics[width=1.0\textwidth]{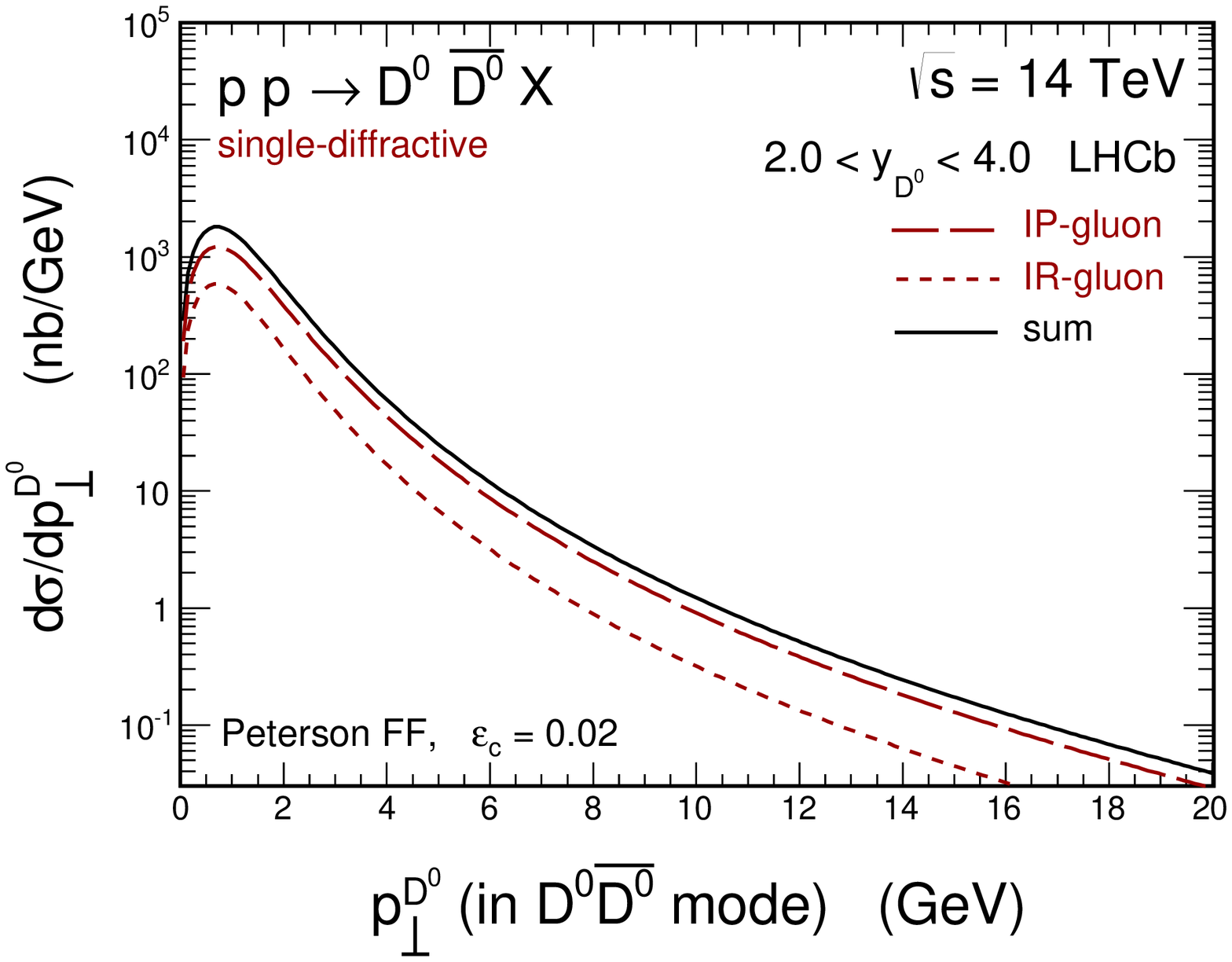}}
\end{minipage}
\hspace{0.5cm}
\begin{minipage}{0.47\textwidth}
 \centerline{\includegraphics[width=1.0\textwidth]{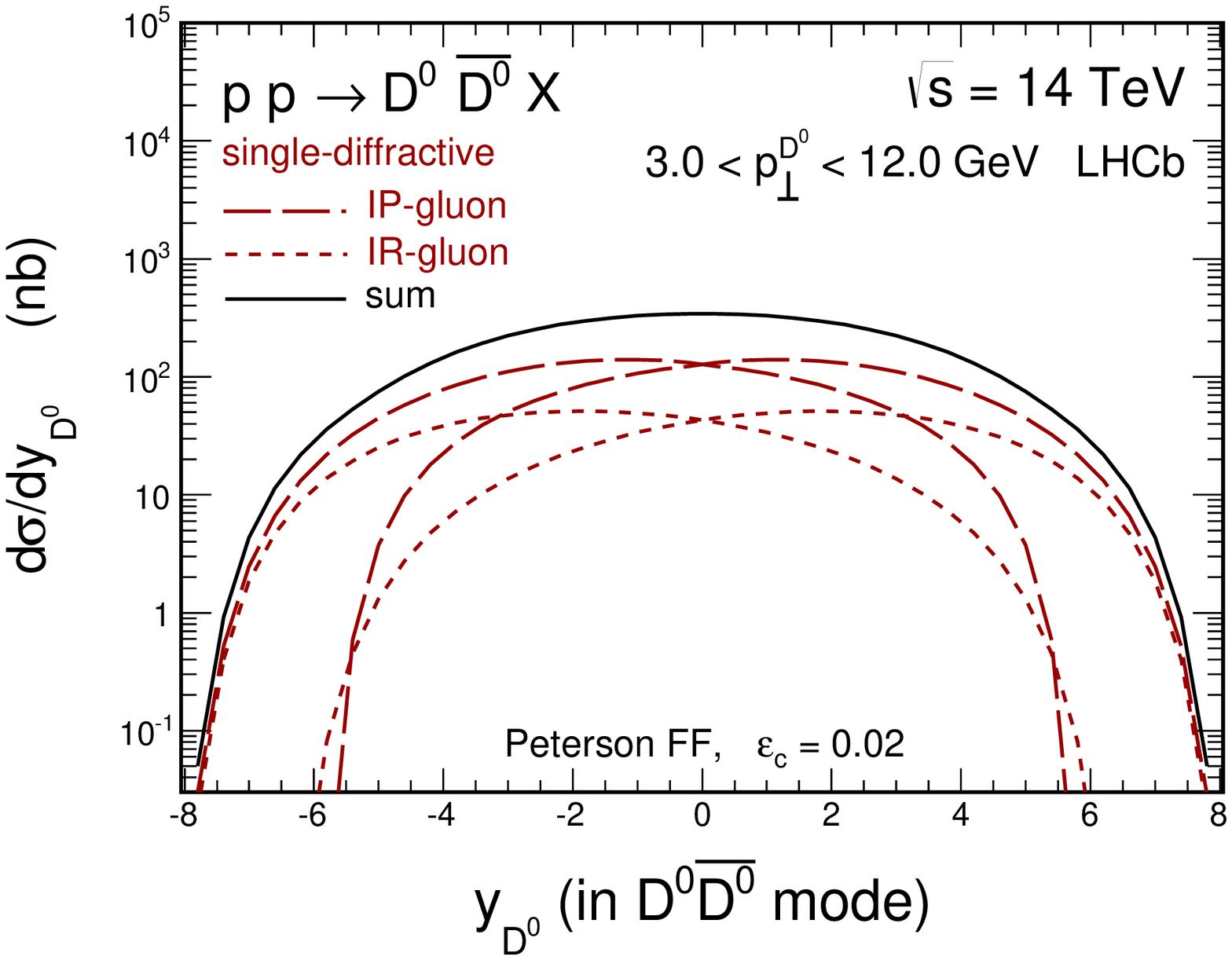}}
\end{minipage}
   \caption{
\small Transverse momentum (left) and rapidity (right) distributions of $D^{0}$ meson within the LHCb acceptance provided that $\overline{D^{0}}$ was registered too, for the single-diffractive mechanisms at $\sqrt{s}=14$ TeV. Components of the pomeron-gluon (and gluon-pomeron) (long-dashed line) and the reggeon-gluon (and gluon-reggeon) (short-dashed line) contributions are shown separately.
}
 \label{fig:pt-SD-DDbar}
\end{figure}

\begin{figure}[!h]
\begin{minipage}{0.47\textwidth}
 \centerline{\includegraphics[width=1.0\textwidth]{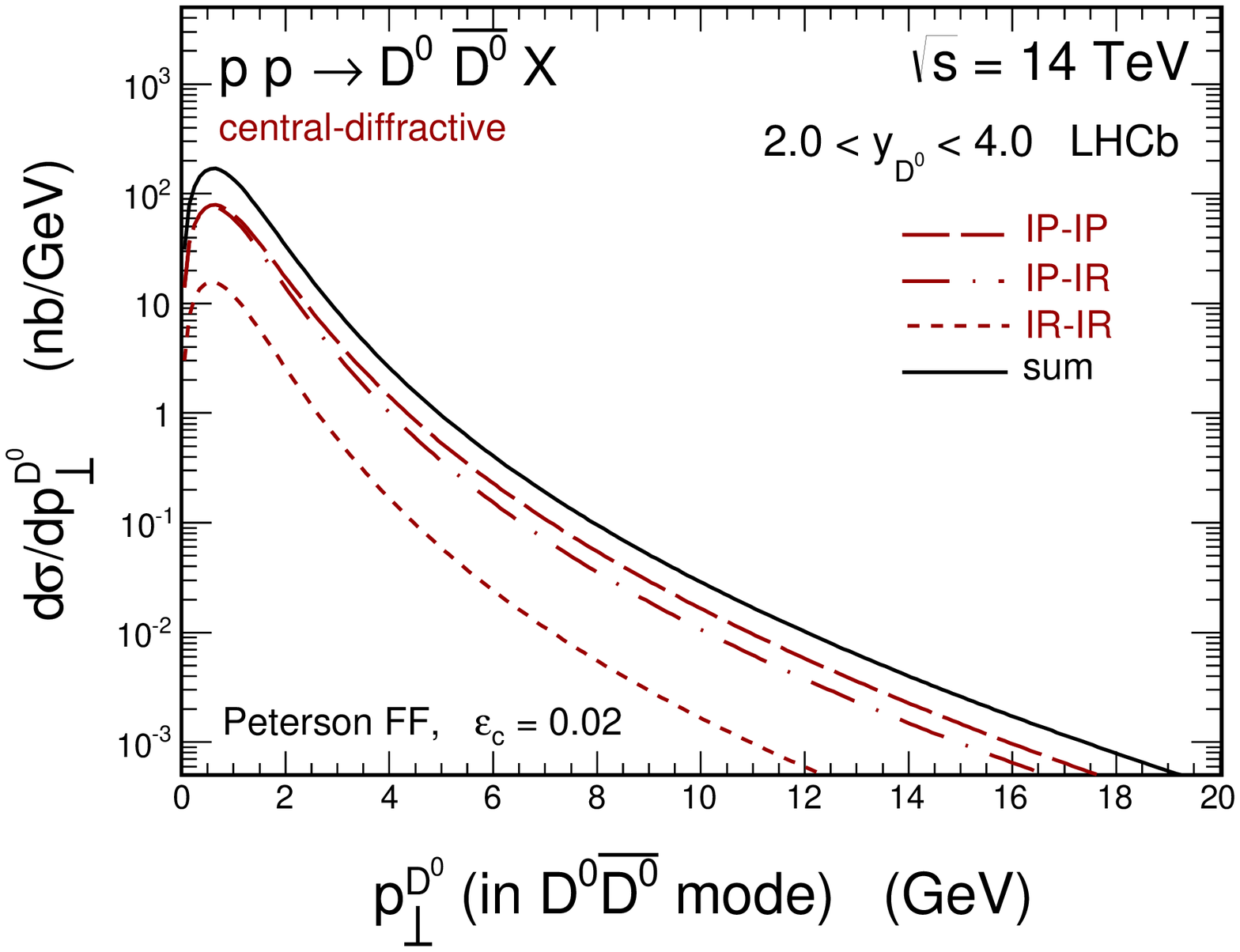}}
\end{minipage}
\hspace{0.5cm}
\begin{minipage}{0.47\textwidth}
 \centerline{\includegraphics[width=1.0\textwidth]{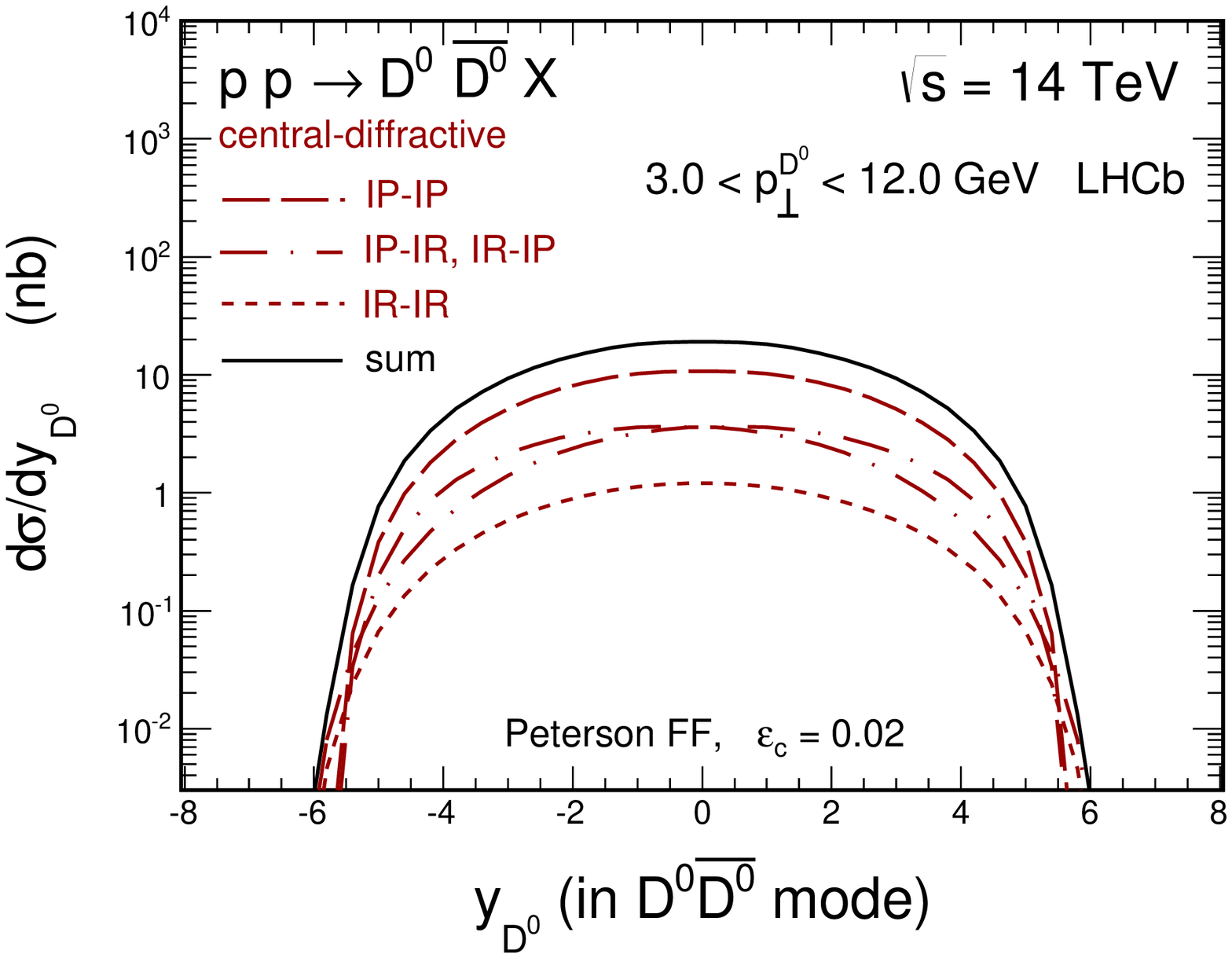}}
\end{minipage}
   \caption{
\small Transverse momentum (left) and rapidity (right) distributions of $D^{0}$ meson within the LHCb acceptance provided that $\overline{D^{0}}$ was registered too, for the central-diffractive mechanism at $\sqrt{s}=14$ TeV. Components of the pomeron-pomeron, pomeron-reggeon, reggeon-pomeron and reggeon-reggeon contributions are shown separately.
}
 \label{fig:pt-CD-DDbar}
\end{figure}

\begin{figure}[!h]
\begin{minipage}{0.31\textwidth}
 \centerline{\includegraphics[width=1.0\textwidth]{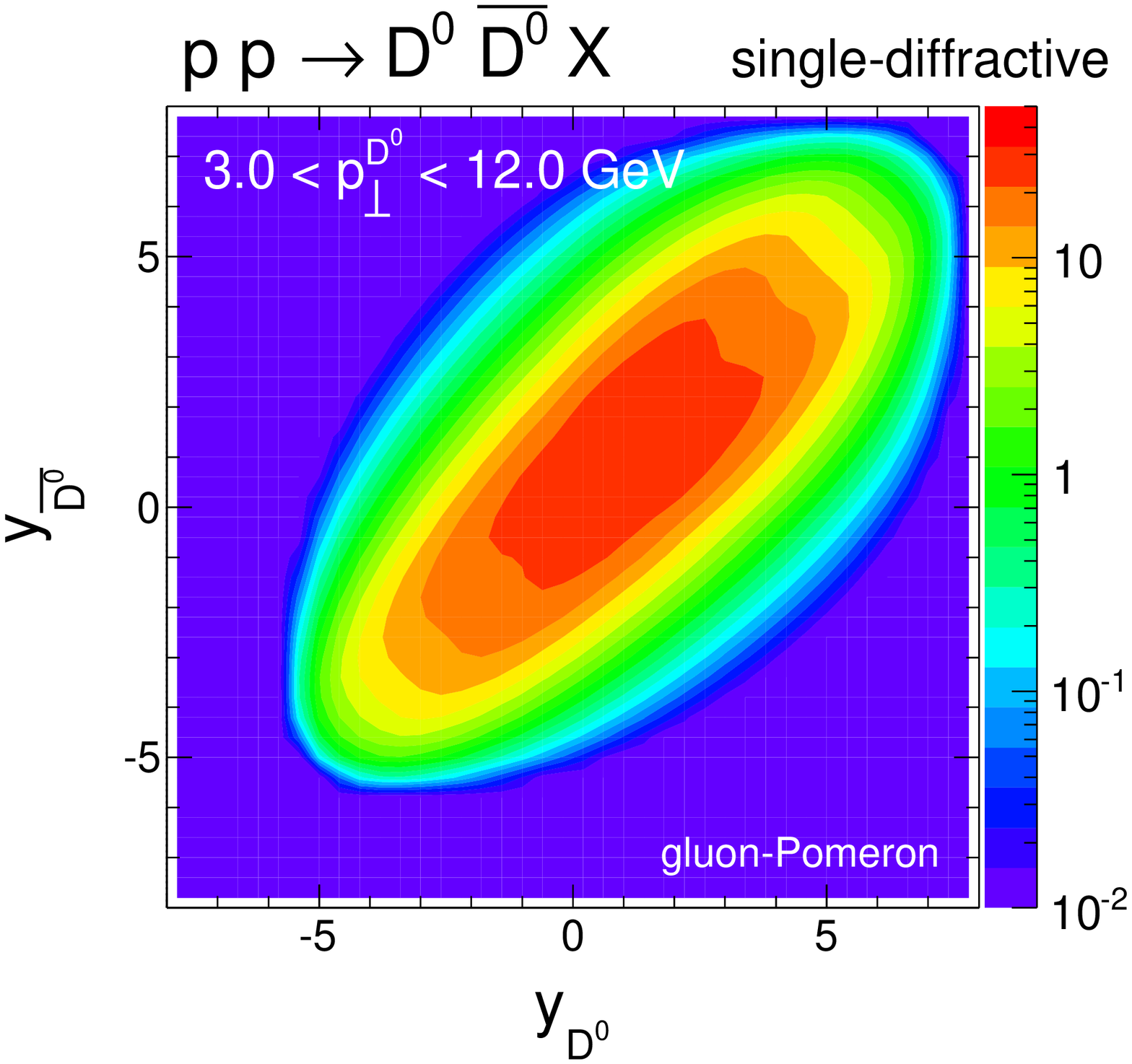}}
\end{minipage}
\hspace{0.2cm}
\begin{minipage}{0.31\textwidth}
 \centerline{\includegraphics[width=1.0\textwidth]{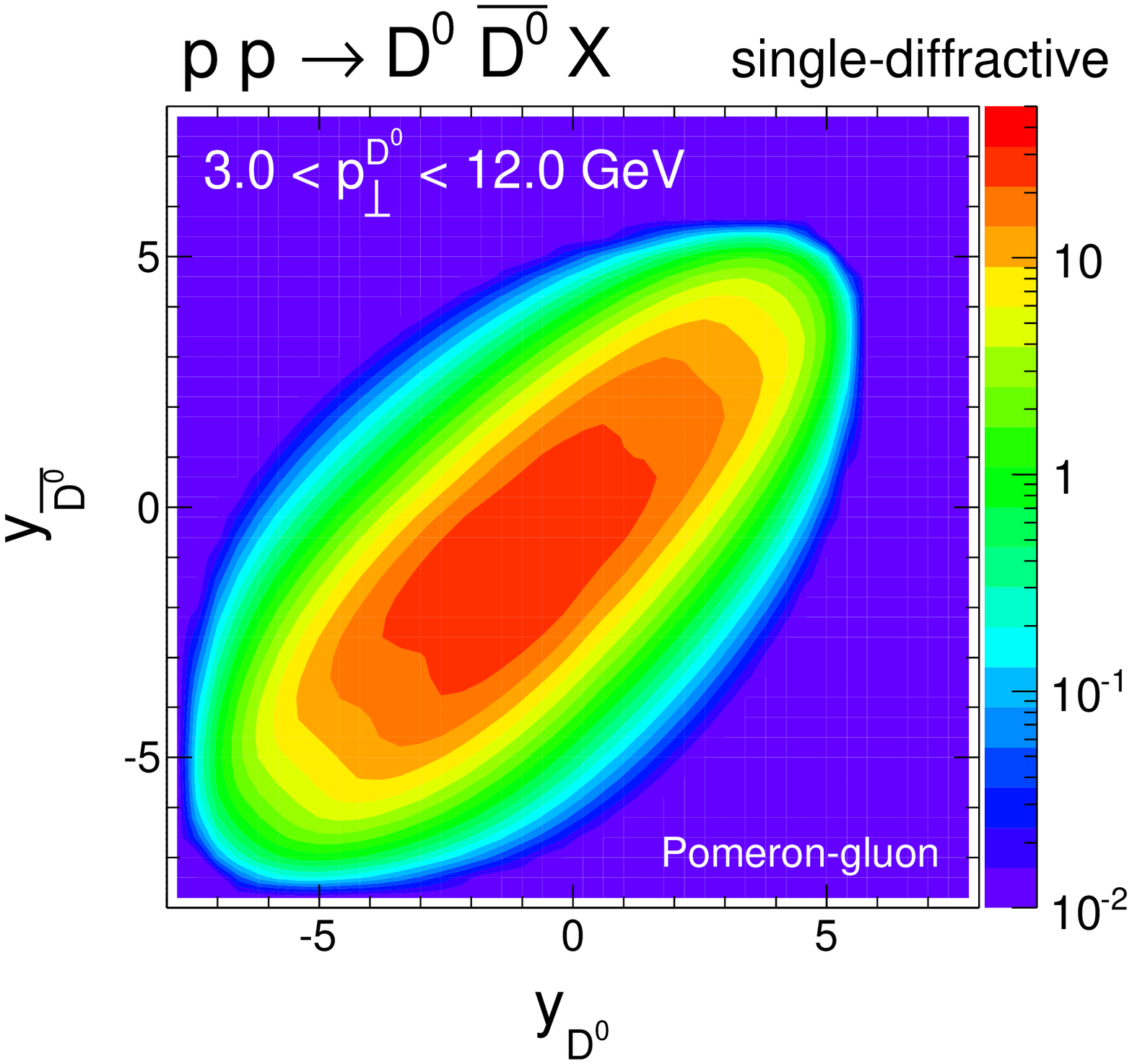}}
\end{minipage}
\hspace{0.2cm}
\begin{minipage}{0.31\textwidth}
 \centerline{\includegraphics[width=1.0\textwidth]{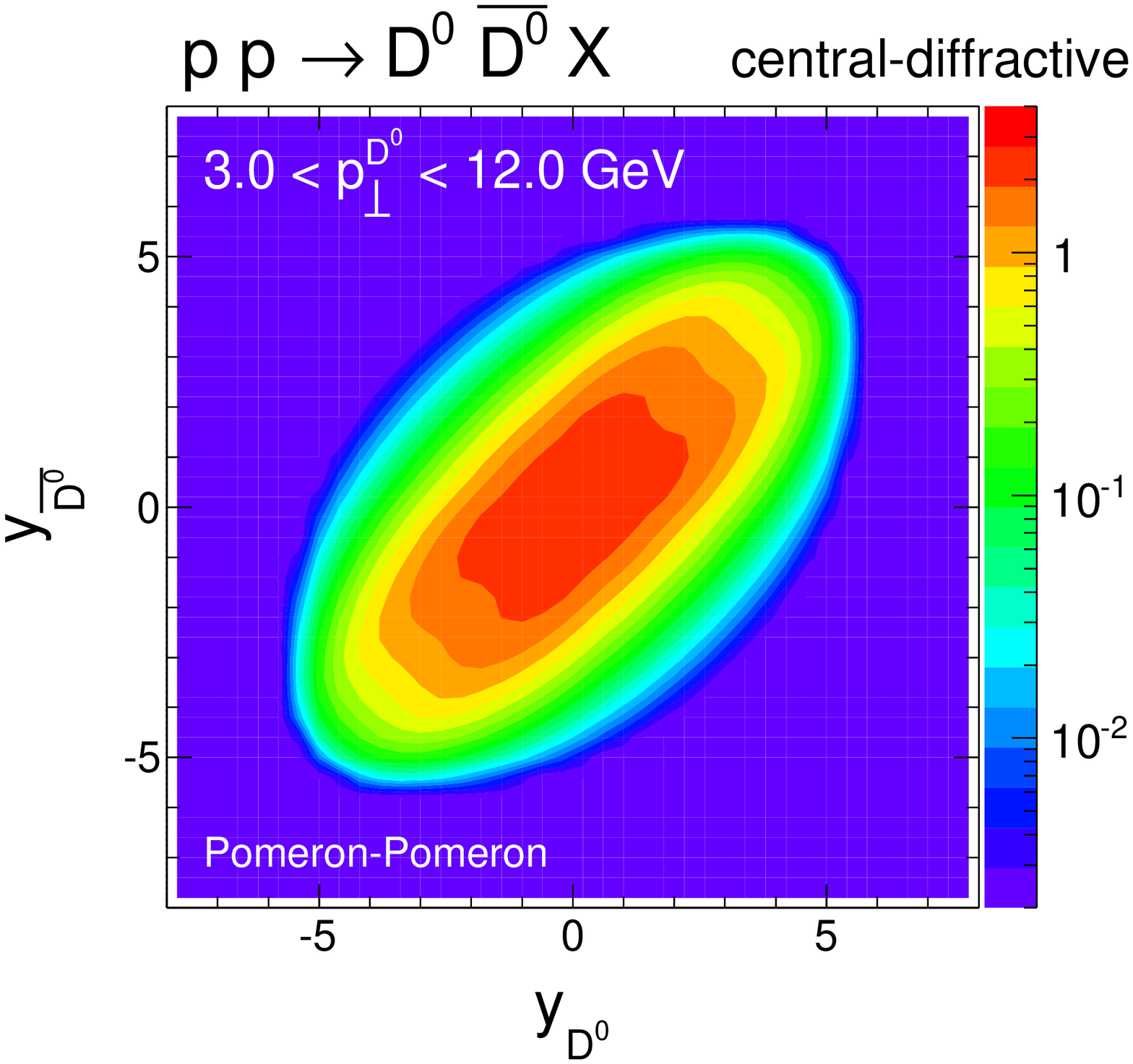}}
\end{minipage}\\

\begin{minipage}{0.31\textwidth}
 \centerline{\includegraphics[width=1.0\textwidth]{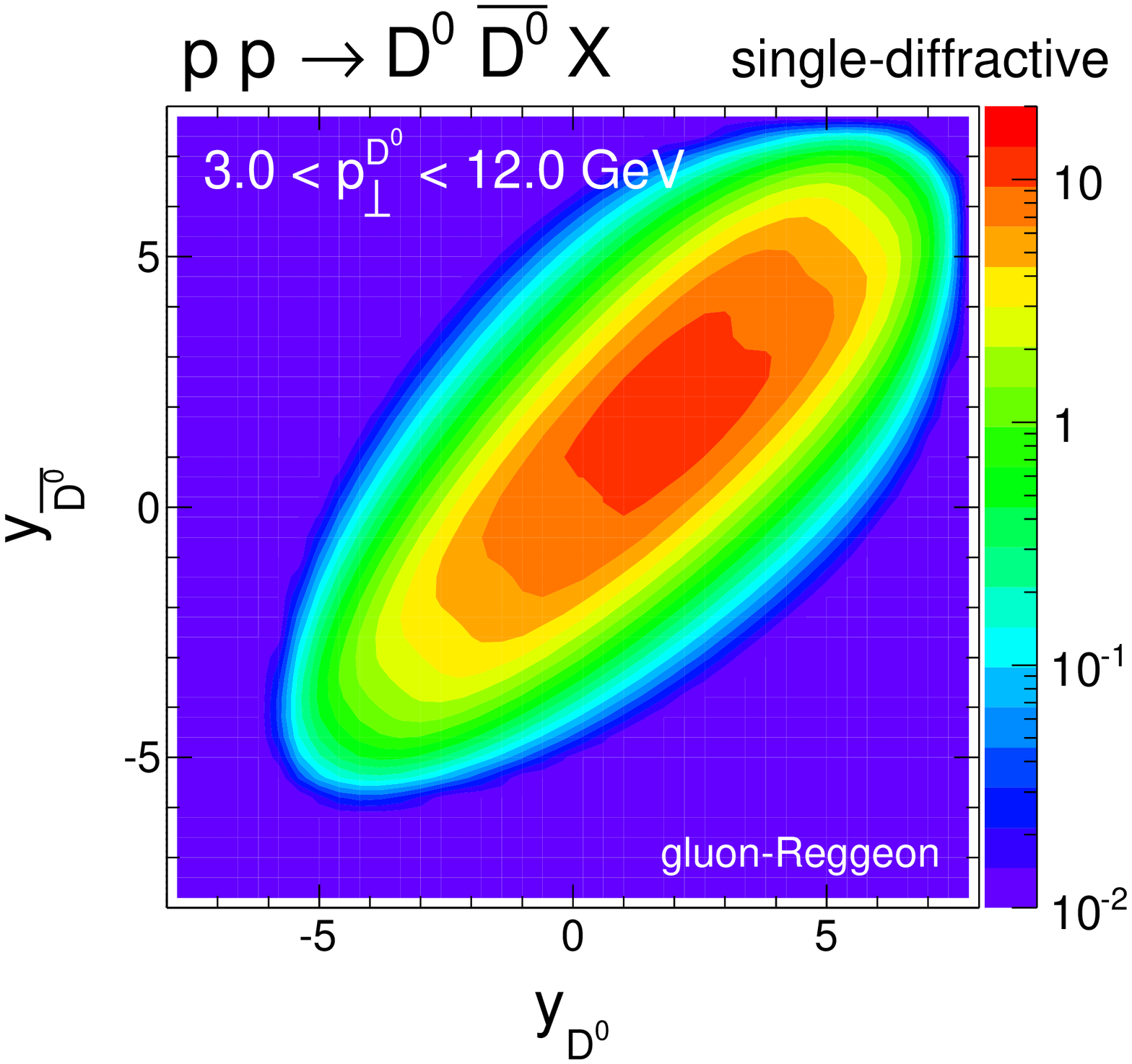}}
\end{minipage}
\hspace{0.2cm}
\begin{minipage}{0.31\textwidth}
 \centerline{\includegraphics[width=1.0\textwidth]{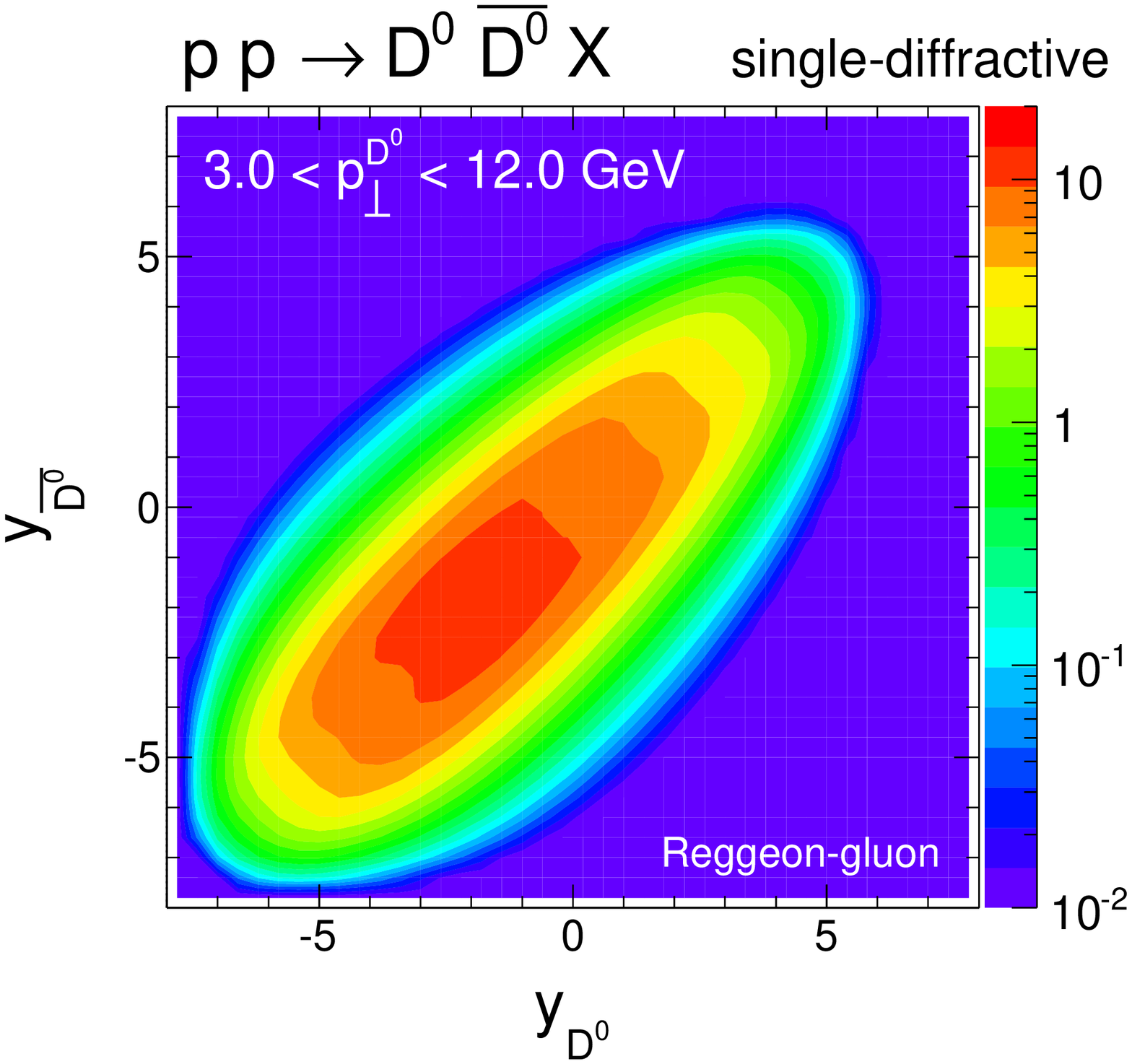}}
\end{minipage}
\hspace{0.2cm}
\begin{minipage}{0.31\textwidth}
 \centerline{\includegraphics[width=1.0\textwidth]{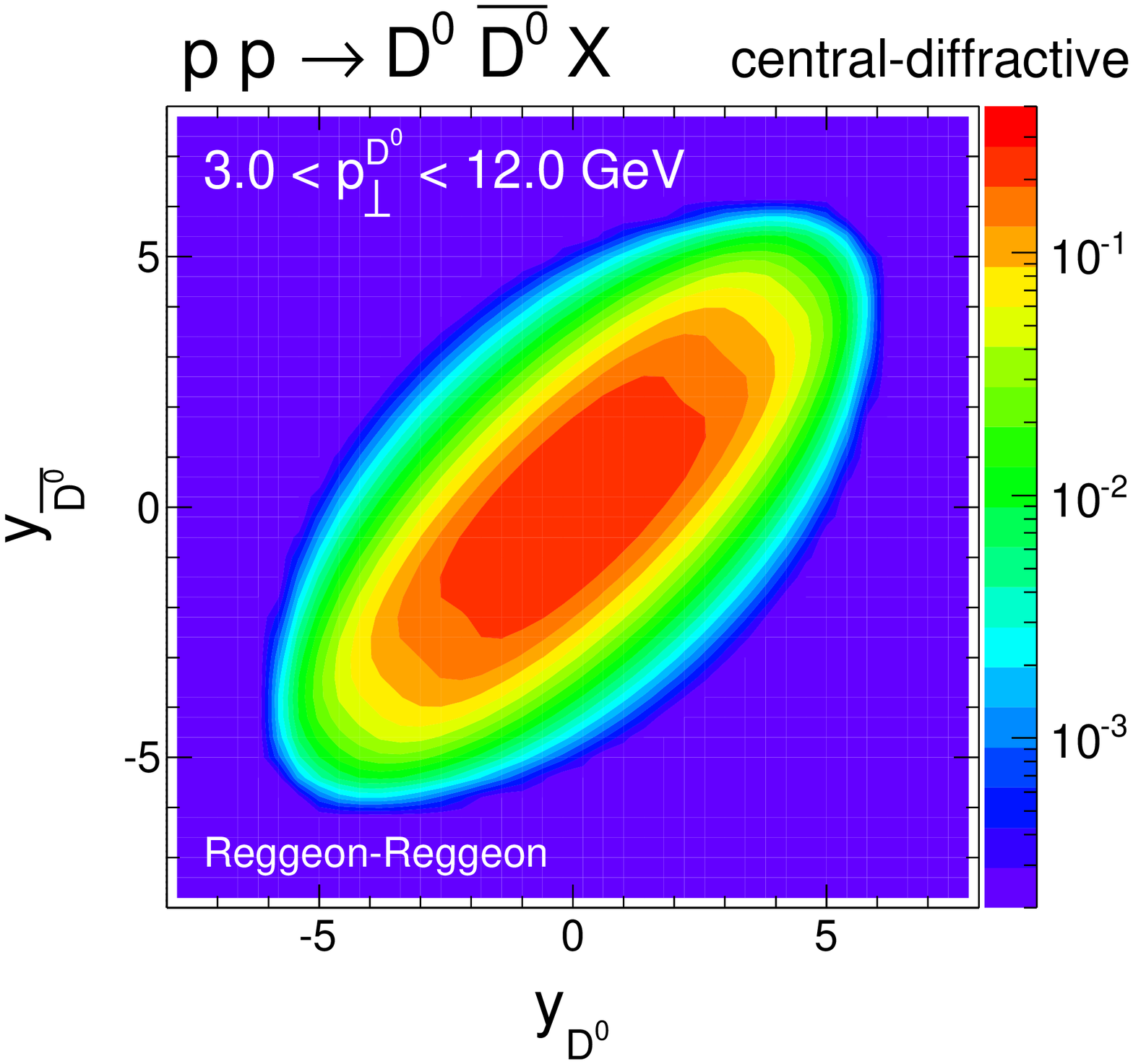}}
\end{minipage}

   \caption{
\small Double differential cross sections as a function of $D^{0}$ and $\overline{D^{0}}$ rapidities within the LHCb detector acceptance for single- (left and middle panels) and central-diffractive (right panels) production at $\sqrt{s}=14$ TeV. The top and bottom panels correspond to the pomeron and reggeon exchange mechanisms respectively. 
}
 \label{fig:2dim-y1y2}
\end{figure}

Figures~\ref{fig:pt-SD-DDbar} and~\ref{fig:pt-CD-DDbar} show transverse momentum (left panels) and rapidity (right panels) distributions of $D^{0}$ (or $\overline{D^{0}}$) meson at $\sqrt{s}=14$ TeV within the LHCb acceptance in the case of $D^{0}\overline{D^{0}}$ pair production for single- and central-diffractive mechanisms, respectively. The graphical representation of pomeron- and reggeon-exchange contributions is the same as in the previous figures.
The rapidity distributions for pomeron-gluon (or reggeon-gluon) and gluon-pomeron (or gluon-reggeon) mechanisms in the single-diffractive case are shifted
to forward and backward rapidities, respectively. Since the rapidity acceptance of the LHCb detector is not symmetric in rapidity and covers only forward region $2<y_{D^{0}}<4$ these both single-diffractive mechanisms contribute to the $D^{0}\overline{D^{0}}$ pair diffractive cross section in a quite different way. The situation is shown in more detail in Fig.~\ref{fig:2dim-y1y2} where the rapidity correlations between $D^{0}$ and $\overline{D^{0}}$ meson are depicted.  
In all the considered cases these distributions show some correlation along the diagonal. Clearly some shifts of the distributions for the single-diffractive
mechanism can be seen, in contrast to the central-diffractive one. 

\section{Conclusions}
Although there was a lot of theoretical activity in calculating
diffractive production of different objects 
(gauge bosons ($W$, $Z$), jets or dijets, Higgs boson, pairs of gauge 
bosons ($W^+ W^-$) in proton-proton or proton-antiproton collisions,
almost no detailed experimental studies were performed and presented
in the literature.
Such a study would be interesting and important in order to understand
mechanism of diffractive production. This is partly so as many reactions
considered so far have rather very small cross section.
So far there is no common agreement on what is underlying mechanism
of diffractive production. 
Since the underlying dynamics is of nonperturbative nature any
detailed studies would be therefore very helpful to shed new light 
on the problem.

In the present paper we discuss in more detail single-
and central- diffractive production of charm and bottom
quark-antiquark pairs as well as open charmed and bottom mesons.
The corresponding cross sections are rather large.

In the present study we have limited ourselves to the most popular
Ingelman-Schlein model of resolved pomeron and reggeon. 
Although there is no experimental proof for the model and its underlying
dynamics it has advantage it was used to describe many diffractive 
processes at HERA. In the purely hadronic 
processes considered in the present paper it must be supplemented 
by including absorption effects due to nonperturbative interaction 
of hadrons (protons).

In our approach we use diffractive parton distribution in the proton
obtained at HERA from the analysis of diffractive structure function
of the proton and diffractive production of jets. Both pomeron
and reggeon contributions are considered here.

First we have calculated cross sections for $c \bar c$ and $b \bar b$
production in single and central production. Several quark-level
differential distributions are shown and discussed. We have compared
pomeron and reggeon contributions for the first time.

In order to make predictions which could be compared with future
experimental data in the next step we have included hadronization to
charmed ($D$) and bottom ($B$) mesons using a practical method
of hadronization functions known for other processes.
We have shown several inclusive differential distributions for the
mesons as well as correlations of $D$ and $\bar D$ mesons.
In these calculations we have included detector acceptance
of the ATLAS, CMS and LHCb collaboration experiments.

The production of charmed mesons is extremely interesting
because of the cross section of the order of a few microbarns for ATLAS 
and CMS and of the order of tens of microbarns for the LHCb acceptance.
We have shown that the pomeron contribution is much larger than
the subleading reggeon contribution.
Especially the LHCb main detector supplemented with
VELO (VErtex LOcator) micro-strip silicon detectors
installed already in Run I and so-called HERSCHEL 
(High Rapidity Shower Counters for LHCb)
apparatus to be installed in Run II 
could be used to measure $D$ mesons (main detector) and define rapidity 
gap nececcesary for diffractive production (VELO and/or HERSCHEL).
On the other hand ATLAS and CMS collaboration could use
ALFA and TOTEM detectors to measure forward protons. Then different
additional differential distributions are possible.


\vspace{1cm}

{\bf Acknowledgments}
We are indebted to Guy Wilkinson from the LHCb collaboration
for discussion of the LHCb possibilities and Wolfgang Sch\"afer
for several interesting discussions of diffractive processes
and heavy quark production in particular.
We wish to thank Rafa{\l} Staszewski and Maciej Trzebi\'nski
for continues interest in our studies.
This work was partially supported by the Polish
NCN grant DEC-2013/09/D/ST2/03724.


\end{document}